\documentclass[aps,pre,floats,floatfix,superscriptaddress,nofootinbib,twocolumn]{revtex4}
\usepackage{bbm}
\usepackage{amsmath,amssymb,amsthm}
\usepackage{xcolor}
\usepackage{graphicx}
\usepackage{caption}
\usepackage{makecell}
\usepackage{url}
\usepackage{answers}
\usepackage{array}
\usepackage{multirow}
\usepackage[mathscr]{euscript}
\usepackage{eufrak}
\usepackage{calligra}

\definecolor{link-color}{cmyk}{0.8 ,  0.3 ,  0. , 0}
\usepackage[breaklinks,
colorlinks=true,%
linkcolor   = {link-color},%
citecolor   = {link-color},%
urlcolor    = {link-color}]{hyperref}

\newcommand{\ZZZ}[2]{}

\newcommand{\kb}{}

\newcommand{\rr}{\mathfrak{r}}
\newcommand{\R}{\mathcal{R}}

\newcommand{\m}{\mathfrak{m}}
\newcommand{\al}{\textsl{a}}

\begin{document}

	\title{Random hyperbolic graphs in $d+1$ dimensions}
	
	\author{Gabriel Budel}
	\affiliation{Faculty of Electrical Engineering, Mathematics and Computer Science, Delft University of Technology, 2628 CD, Delft, Netherlands}
	\author{Maksim Kitsak}
	\affiliation{Faculty of Electrical Engineering, Mathematics and Computer Science, Delft University of Technology, 2628 CD, Delft, Netherlands}
	\author{Rodrigo Aldecoa}
	\affiliation{Department of Physics, Northeastern
		University, Boston, Massachusetts 02115, USA}
	\affiliation{Network Science Institute, Northeastern University, Boston, Massachusetts 02115, USA}
	\author{Konstantin Zuev}
	\affiliation{Department of Computing and Mathematical Sciences, California Institute of Technology, California 91125, USA}
	
	\author{Dmitri Krioukov}
	\affiliation{Department of Physics, Department of Mathematics, Department of Electrical\&Computer Engineering,
		Northeastern University, Boston, Massachusetts 02115, USA}
	\affiliation{Network Science Institute, Northeastern University, Boston, Massachusetts 02115, USA}

	\begin{abstract}
		We consider random hyperbolic graphs in hyperbolic spaces of any dimension $d+1\geq 2$. We present a rescaling of model parameters that casts the random hyperbolic graph model of any dimension to a unified mathematical framework, leaving the degree distribution invariant with respect to the dimension. Unlike the degree distribution, clustering does depend on the dimension, decreasing to 0 at $d \rightarrow \infty$. We analyze all of the other limiting regimes of the model, and we release a software package that generates random hyperbolic graphs and their limits in hyperbolic spaces of any dimension.
	\end{abstract}
	\maketitle
	
	\section{Introduction}
	
	Random hyperbolic graphs (RHGs)~\cite{Krioukov2009,Krioukov2010hyperbolic} are a latent space network model~\cite{mcfarland1973social,hoff2002latent}, in which the latent space is the hyperbolic plane~$\mathbb{H}^{2}$: nodes are random points on the plane, while connections between nodes are established with distance-dependent probabilities. RHGs reproduce many structural properties of real networks including sparsity, self-similarity, power-law degree distribution, strong clustering, small worldness, and community structure~\cite{Serrano2008,Krioukov2010hyperbolic,papadopoulos2012popularity,zuev2015emergence,zheng2021scaling}. They are also exponential random graphs with just two sufficient statistics---the number of links and the sum of their hyperbolic lengths~\cite{boguna2020small}. Using the RHG as a null model, one can map real networks to hyperbolic spaces~\cite{Boguna2010sustaining,kitsak2020link,Perez2019mercator}, the applications of which include routing and navigation~\cite{Boguna2010sustaining,boguna2008,Gulyas2015,voitalov2017geohyperbolic,Ortiz2017navigability,Garcia2018multiscale,Muscoloni2019navigability}, link prediction~\cite{Serrano2012uncovering,Papadopoulos2015network1,Papadopoulos2015network,Muscoloni2017machine,Alessandro2018leveraging,Muscoloni2018minimum,perez2020precision,kitsak2020link}, network scaling~\cite{Serrano2008,Garcia2018multiscale,zheng2021scaling}, semantic analysis~\cite{nickel2017poincare,nickel2018learning,dhingra2018embedding,ifrea2019poincare}, inference of shortest paths~\cite{kitsak2023finding}, and many others~\cite{boguna2020network}.
	
	While two-dimensional RHGs have been studied extensively in the literature, their higher-dimensional generalizations have not received a lot of attention. Here, we aim to fill in this gap by offering a systematic analysis of the RHG model in a hyperbolic space~$\mathbb{H}^{d+1}$ of arbitrary dimensionality $d + 1 \geq 2$. Apart from purely academic interest, our work is inspired by several practical questions. Hyperbolic spaces expand exponentially for any dimension $d + 1 \geq 2$. Thus, intuitively, RHGs should have similar topological properties regardless of the dimensionality of their latent hyperbolic spaces. We aim to verify this intuition here. Second, the dimensionality of the latent space has been shown to affect the accuracy of graph embedding tasks~\cite{gu2021principled}. Finally, a recent study suggests that the dimensionality of the hyperbolic space allows one to generate more realistic and diverse community structures~\cite{desy2023dimension}.

	Spatial graphs in hyperbolic spaces comprise an area of active research, and many results were obtained recently. RHGs are equivalent to geometric inhomogeneous random graphs (GIRGs), as mentioned in~\cite{Krioukov2010hyperbolic} and formalized in~\cite{bringmann2019geometric}. This GIRG formulation is followed in~\cite{Boguna2009random, boguna2020small,almagro2022detecting,friedrich2023cliques}, where the small-world, clustering, and other properties of GIRGs are analyzed for any dimension~$d$. The work by W. Yang and D. Rideout~\cite{yang2020high} contains rigorous mathematical proofs for degree distributions and degree correlations of the high-dimensional RHG model. The related popularity-similarity optimization (PSO) model has recently been extended to arbitrary dimensionality $d + 1 \geq 2$ in \cite{kovacs2022generalised}.  Whereas RHGs are a static network model, the $(d+1)$-dimensional PSO model is a growing network model in hyperbolic space that possesses similar structural properties.

	Here, we conduct a systematic analysis of the structural properties of RHGs and their limiting regimes. In Sec.~\ref{sec:geometric-ergs}, we define the RHG model in hyperbolic spaces~$\mathbb{H}^{d+1}$ of any dimension $d+1\geq2$. We present a rescaling of the model parameters that renders the degree distribution invariant with respect to~$d$, Sec.~\ref{sec:RHG_pk}, focusing on the three connectivity regimes---cold, critical, and hot---in the model, Sec.~\ref{sec:RHG_connectivity}. In Sec.~\ref{sec:limiting}, we analyze the limiting regimes of the model when its parameters tend to their extreme values. These regimes are Erd\H{o}s-R\'{e}nyi random graphs, the configuration model, and (soft) random geometric graphs on spheres. Sec.~\ref{sec:code} introduces our software package that generates RHGs and their limits for any~$d$, generalizing the $d=1$ generator in~\cite{aldecoa2015hyperbolic}. The concluding remarks are in Sec.~\ref{sec:summary}.

	\section{Random Hyperbolic Graph Model in  $d+1$ dimensions}\label{sec:geometric-ergs}
	
	A random hyperbolic graph (RHG)  in the $(d+1)$-dimensional hyperbolic space $\mathbb{H}^{d+1}$ is defined as follows. Every node $i$ of the RHG corresponds to a point $\mathbf{x}_{i}$ in $\mathbb{H}^{d+1}$. Points are selected uniformly at random with a pdf $\rho(\mathbf{x})$, which is prescribed by the model. Connections between node pairs in the RHG are established with independent probabilities $p:R^{+} \to [0,1]$ that are functions of the distances in $\mathbb{H}^{d+1}$. The probability $p_{ij}$ of a link between nodes $i$ and $j$ is $p_{ij} = p\left(d_{\mathbb{H}^{d+1}}\left(\mathbf{x}_{i}, \mathbf{x}_{j}\right) \right)$, where $d_{\mathbb{H}^{d+1}}
		\left(\mathbf{x}_{i}, \mathbf{x}_{j}\right)$ is the distance between points $\mathbf{x}_{i}$ and $\mathbf{x}_{j}$ in~$\mathbb{H}^{d+1}$. 
		
		The RHG model, as a result, is fully defined by its latent space $\mathbb{H}^{d+1}$, node pdf $\rho(\mathbf{x})$, and the connection probability function $p(d)$. 
		
		To justify our choices for $\rho(\mathbf{x})$ and $p(d)$, we first recall the definition of the $d$-dimensional hyperbolic space and its basic geometric properties. To this end, we consider the upper sheet of the $(d+1)$-dimensional hyperboloid of curvature $K=-\zeta^{2}$,
	\begin{equation}
		x_0^{2} - x_1^{2} - ... - x_{d+1}^{2} = \frac{1}{\zeta^{2}},~x_0 > 0,
	\end{equation}
	in the $(d+2)$-dimensional Minkowski space with metric
	\begin{equation}
		{\rm d} s^{2}  = -{\rm d} x_0^{2} + {\rm d} x_{1}^{2} + ... + {\rm d} x_{d+1}^{2}.
		\label{eq:minkowski_metric}
	\end{equation}
	The spherical coordinate system on the hyperboloid $\left(r, \theta_1, ...,  \theta_d \right)$ is defined by
	\begin{align}
		\label{eq:hyperboloid_coordinates}
		x_0 =& \frac{1}{\zeta} \cosh \zeta r, \nonumber \\
		x_1 =& \frac{1}{\zeta} \sinh \zeta r \cos \theta_1, \nonumber\\
		x_2 =& \frac{1}{\zeta} \sinh \zeta r \sin \theta_1 \cos \theta_2,\\
		\vdots\nonumber \\
		x_d =& \frac{1}{\zeta} \sinh \zeta r \sin \theta_1 ... \sin \theta_{d-1} \cos \theta_{d},\nonumber\\
		x_{d+1} =& \frac{1}{\zeta} \sinh \zeta r \sin \theta_1 ... \sin \theta_{d-1} \sin \theta_{d} \nonumber,
	\end{align}
	where $r > 0$ is the radial coordinate and $\left(\theta_1,...,\theta_d\right)$ are the standard angular coordinates on the unit $d$-dimensional sphere $\mathbb{S}^{d}$.
	
	The coordinate transformation in (\ref{eq:hyperboloid_coordinates}) yields the spherical coordinate metric in the $(d+1)$-dimensional hyperbolic space $\mathbb{H}^{d+1}$
	\begin{align}
		{\rm d} s^{2} &= {\rm d} r^{2} + {1 \over \zeta^{2} }\sinh^{2} \left({ \zeta r} \right) {\rm d} \Omega_{d}^{2},\\
		{\rm d} \Omega_{d}^{2} &=  {\rm d} \theta_{1}^{2} + {\rm sin}^{2} (\theta_{1}){\rm d}
		\theta_{2}^{2}+\dots\nonumber \\
		&\quad+ {\rm sin}^{2} (\theta_{1}) \dots {\rm sin}^{2} (\theta_{d-1}) {\rm d}
		\theta_{d}^{2}
		\label{eq:Hd_metric},
	\end{align}
	resulting in the volume element in  $\mathbb{H}^{d+1}$:
	\begin{equation}
		{\rm d}V =  \left({\frac{1} {\zeta} }\sinh {\zeta r}\right)^{d}  {\rm d}r \prod_{k\,=\,1}^{d} \sin^{d-k}
		(\theta_{k}) {\rm d} \theta_{k}.
	\end{equation}

	The distance between two points $i$ and $j$ in $\mathbb{H}^{d+1}$ is given by the hyperbolic law of
	cosines:
	\begin{equation}
		\label{eq:cosh-law} \cosh \zeta d_{ij} = \cosh \zeta r_i \cosh \zeta r_j - \sinh \zeta r_i \sinh
		\zeta r_j \cos \Delta \theta_{ij},\\
	\end{equation}
	where $\Delta \theta_{ij}$ is the angle between $i$ and $j$:
	\begin{align}
		&\cos (\Delta \theta_{ij}) = \cos \theta_{i,1} \cos \theta_{j,1}\nonumber\\
		+& \sin \theta_{i,1} \sin \theta_{j,1} \cos
		\theta_{i,2} \cos \theta_{j,2} + ... \nonumber \\
		+&\sin \theta_{i,1} \sin \theta_{j,1}... \sin \theta_{i,d-1} \sin \theta_{j,d-1} \cos \theta_{i,d} \cos
		\theta_{j,d} \nonumber \\
		+& \sin \theta_{i,1} \sin \theta_{j,1}... \sin \theta_{i,d-1} \sin \theta_{j,d-1} \sin \theta_{i,d} \sin
		\theta_{j,d},
		\label{eq:cosine_theta_hd}
	\end{align}
	where $\left(\theta_{i,1},...,\theta_{i,d}\right)$ and $\left(\theta_{j,1},...,\theta_{j,d}\right)$ are the coordinates of the points $i$ and $j$ on $\mathbb{S}^{d}$.

	For sufficiently large $\zeta r_{i}$ and $\zeta r_{j}$ values,  the hyperbolic law of cosines
	in Eq.~(\ref{eq:cosh-law}) is closely approximated by
	\begin{equation}
		\label{eq:cosh-law_approx} d_{ij} \approx r_i + r_j + {2 \over \zeta} \ln \left( \sin (\Delta
		\theta_{ij} /2) \right).
	\end{equation}

	The hyperbolic ball $\mathbb{B}^{d+1}$ of radius $R > 0$ is defined as the set of points with
	\begin{equation}
		r \in [0, R].
		\label{eq:Hd_coords}
	\end{equation}

	Nodes of the RHG are points in $\mathbb{B}^{d+1}$ selected at random with density  $\rho_{\mathbf{x}}(\mathbf{x}) \equiv \rho_{r}(r)\rho_{\theta_1}(\theta_1)...\rho_{\theta_d}(\theta_d)$, where 
	\begin{subequations}
		\label{eq:hd_rho_functions}
		\begin{align}
			\rho_{r}(r) &= \left[\sinh(\alpha r)\right]^{d}/C_{d},\qquad\alpha \geq \zeta/2,
			\label{eq:hd_rho_r}\\
			C_{d} &\equiv \int_{0}^{R} \left[\sinh(\alpha r)\right]^{d} {\rm d} r,
			\label{eq:hd_cd}\\
			\rho_{\theta_k}(\theta_k) &= \left[{\rm sin} (\theta_k) \right]^{d-k}/I_{d,k},\qquad k=1,\ldots,d,
			\label{eq:hd_rho_theta}\\
			I_{d,k} &\equiv \int_{0}^{\pi} \left[\sin (\theta_k) \right]^{d-k} {\rm d} \theta_k \nonumber\\
			&= \sqrt{\pi} \frac{\Gamma[\frac{d-k+1}{2}]}{\Gamma\left[1 + \frac{d-k}{2}\right]},\qquad k<d,
			\label{eq:hd_I_dk}\\
			I_{d,d} &\equiv 2 \pi.
			\label{eq:hd_I_dd}
		\end{align}
	\end{subequations}
	In other words,  nodes are uniformly distributed on unit sphere $\mathbb{S}^{d}$ with respect to their angular coordinates. In the special case of $\alpha = \zeta$ nodes are also uniformly distributed in $\mathbb{B}^{d+1}$.

	Pairs of nodes $i$ and $j$ are connected independently with connection probability
	\begin{equation}
		p_{ij} = p\left(d_{ij}\right)= {1 \over 1 + {\rm
				exp}\left(\zeta \left(d_{ij} - \mu \right)/\kb 2 T\right)} \label{eq:conn_hypergeometric_ERG},
	\end{equation}
	where $\mu>0$ and $T>0$ are model parameters and $d_{ij}$ is the distance between points $i$ and $j$ in $\mathbb{B}^{d+1}$, given by Eq.~(\ref{eq:cosh-law}). We refer to parameters $T$ and $\mu$ as the temperature and the chemical potential, respectively, using the analogy with the Fermi-Dirac statistics. We note that the factors of $2$ and $\zeta$ in Eq.~(\ref{eq:conn_hypergeometric_ERG}) are to agree with the 2-dimensional RHG~\cite{Krioukov2010hyperbolic} that corresponds to $d=1$.
	
\begin{figure*}
		\includegraphics[width=12cm]{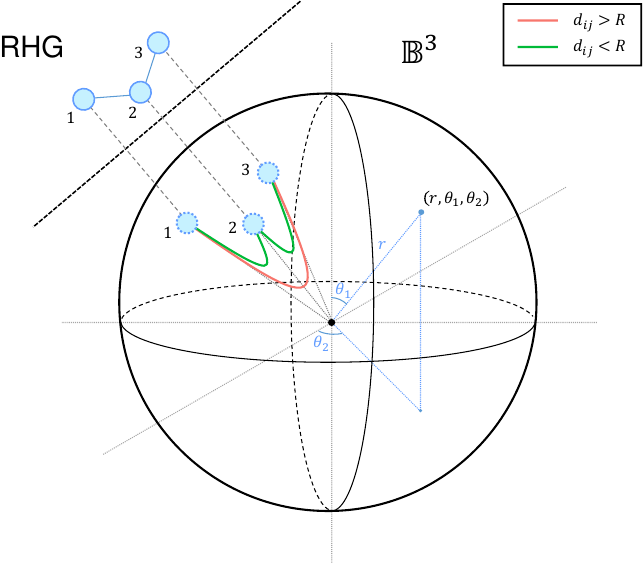}
		\caption{A visualization of an RHG for $d=2$ in $\mathbb{B}^{3}$. The spherical coordinates $(r, \theta_{1}, \theta_{2})$ of a point in $\mathbb{B}^{3}$ are shown. A subgraph of 3 nodes is shown with geodesics between the nodes. Node pairs $(1,2)$ and $(2,3)$ are connected with high probability because $d_{12} < R$ and $d_{23} < R$ (green), while node pair $(1,3)$ is disconnected with high probability because $d_{23} > R$ (red).}
		\label{fig:rhg_fig}
	\end{figure*}

	Thus, the RHG is formed in a three-step network generation process:
	\begin{enumerate}
		\item Randomly select $n$ points in $\mathbb{B}^{d+1}$ with pdf $\rho_{\mathbf{x}}(\mathbf{x})$ in Eq.~(\ref{eq:hd_rho_functions}).
		\item Calculate distances in $\mathbb{B}^{d+1}$ between all node pairs $i$-$j$ using Eq.~(\ref{eq:cosh-law}).
		\item  Connect node pairs $i$-$j$ independently at random with distance-dependent connection probabilities $p_{ij} = p\left(d_{ij}\right)$, prescribed by Eq.~(\ref{eq:conn_hypergeometric_ERG}).
	\end{enumerate}
	Taken together, RHGs in  $\mathbb{B}^{d+1}$ are fully defined by $6$ parameters:
	properties of the hyperbolic ball, $R$  and $\zeta$; number of nodes $n$; radial component $\alpha$ of  the node distribution; chemical potential $\mu$ and temperature $T$. Figure~\ref{fig:rhg_fig} visualizes an RHG for $d=2$.
	
	Only the four parameters $\left(n,\alpha,T,R\right)$ are independent, however. It follows from (\ref{eq:cosh-law}) that $\zeta$ is merely a rescaling parameter for the distances
	$\{d_{ij}\}$ and can be absorbed into the radial coordinates~$r$ by the appropriate rescaling. 
	Chemical potential~$\mu$ controls the expected number of links and the sparsity of the resulting networks. We demonstrate below that the sparsity requirement uniquely determines $R$ and $\mu$ as functions of the network size $n$, $R = R(n)$ and $\mu = \mu(n)$.
	
	\section{Degree distribution and clustering coefficient in the RHG}
	\label{sec:RHG_pk}
	
	The structural properties of the RHG can be computed with the hidden variable formalism, Ref.~\cite{boguna2003class},
	by treating node coordinates as hidden variables.
	
	We begin by calculating the expected degree of node $l$ located at point $\mathbf{x}_{l} = \{r_{l}, \theta_{1}^{l},..., \theta_{d}^{l}\}$:
	\begin{equation}
		\langle k (\mathbf{x}_{l})\rangle = (n-1)\int   {{\rm d} \mathbf{x}_{k} \rho_{\mathbf{x}_{k}}(\mathbf{x}_{k})
			\over 1 + e^{\frac{ \zeta(d_{lk}-\mu)}{2T}}}
		\label{eq:hd_avg_k_preliminary}
	\end{equation}
	
	The symmetry in the angular distribution of points ensures that the expected degree of the node depends
	only on its radial coordinate $r_{l}$ and not on its angular coordinates, $\langle k
	(\mathbf{x}_{l}) \rangle = \langle k (r_l,0,\ldots,0) \rangle \equiv \langle k (r_l) \rangle$. This
	allows us to integrate out the $d$ angular coordinates in Eq.~(\ref{eq:hd_avg_k_preliminary}).

	We also note that the choice for the radial coordinate distribution given by Eq.~(\ref{eq:hd_rho_r}) with $\alpha \geq \zeta/2$ results in most of the
	nodes having large radial coordinates, $r_{i}~\approx~R~\gg~1$. This fact allows us to approximate
	the distances using Eq.~(\ref{eq:cosh-law_approx}):
	\begin{equation}
		\langle k (r_{l}) \rangle \approx \int_{0}^{R}  \int_{0}^{\pi}{  (n-1)  \rho_{r}(r) {\rm d} r \rho_{\theta_{1}}(\theta_{1})
			{\rm d} \theta_{1} \over 1 + \left[ e^{\zeta\left( r + r_{l} - \mu\right)} \left(\sin\left(
			{\theta_{1} \over 2} \right)\right)^{2}\right]^{1 \over 2T}}
		\label{eq:hd_avgk1}.
	\end{equation}

	To further simplify the calculations, we perform the following change of variables and parameters:
	\begin{subequations}
		\label{eq:rescaling}
		\begin{align}
			\{\rr,\rr_l,\R, \m \} &=  {d\zeta \over 2}\{r,r_l,R,\mu\}, 
			\label{eq:rescaling1}\\
			\tau &= d T,
			\label{eq:rescaling2}\\
			\al &= 2\alpha/\zeta,
			\label{eq:rescaling3}
		\end{align}
	\end{subequations}
	where Eq.~(\ref{eq:rescaling1}) corresponds to four equations, each corresponding to one variable or parameter in the brackets.
	
	In terms of the rescaled variables, the connection probability is
	\begin{equation}
		p_{ij}\approx {1 \over 1 + e^{\frac{\rr_i + \rr_j - \m}{\tau} }\left[{\rm sin} \left( \frac{\Delta \theta_{ij}}{2}\right)\right]^\frac{d}{\tau}}, 
		\label{eq:conn_rescaled_ERG}
	\end{equation}
	while Eq.~(\ref{eq:hd_avgk1}) now reads
	\begin{equation}
		\langle k (\rr_{l}) \rangle \approx \int_{0}^{\R}  \int_{0}^{ \pi}{  (n-1)  \rho_{\rr} (\rr) {\rm d} \rr
			\rho_{\theta_{1}}(\theta_{1}){\rm d} \theta_{1} \over 1 +  e^{\frac{ \rr + \rr_{l} - \m }{\tau}} \left[ {\rm sin} \left(\frac{\theta_{1}}{2} \right) \right]^{\frac{d}{\tau}}}
		\label{eq:hd_avgk2},
	\end{equation}
	where the rescaled density
		\begin{equation}
			\rho_{\rr}(\rr) = \frac{\al}{C_{d}}\left[\sinh\left(\dfrac{\al\rr}{d}\right)\right]^{d},\quad0\leq\rr\leq\R, \label{eq:hd_rho_distr_rescaled}
		\end{equation}
		with $C_{d}$ as defined in Eq.~(\ref{eq:hd_cd}). In our analysis below, we operate with $\R \gg 1$ and $\al > 1$. In this case most nodes are characterized by large $\rr$ values and the density $\rho_{\rr}(\rr)$ is approximated as 
		\begin{equation}
			\rho_{\rr}(\rr) \approx \al e^{ \al\left(\rr - \R \right)}.
			\label{eq:hd_rho_distr_rescaled_approx}
		\end{equation}
		The density $\rho_{\rr}(\rr)$ in Eq.~(\ref{eq:hd_rho_distr_rescaled_approx}) is normalized only approximately under the assumptions of $\R \gg 1$ and $\al > 1$: 
		\begin{equation}
			\int_{0}^{\R} \rho_{\rr}(\rr) d\rr \approx 1 - e^{-\al \R}.
		\end{equation}
	
	The expected degree of the graph is given by
	\begin{equation}
		\langle k \rangle = \int_{0}^{\R} {\rm d} \rr \rho_{\rr}(\rr) \langle k (\rr) \rangle
		\label{eq:hd_avgk3},
	\end{equation}
	and  the degree distribution of the RHG can be expressed as
	\begin{equation}
		P(k) =  \int_{0}^{\R} {\rm d} \rr \rho_{\rr}(\rr) P(k|\rr),
	\end{equation}
	where $P(k|\rr)$ is a conditional probability that  a node with radial coordinate $\rr$ has exactly $k$ connections.
	
	In the case of sparse graphs $P(k|\rr)$ is closely approximated by the Poisson distribution:
	\begin{equation}
		P(k|\rr) \approx {1 \over k!} e^{-\langle k (\rr)\rangle } \left[\langle k (\rr)\rangle \right]
		^{k},
		\label{eq:hd_cond_pk}
	\end{equation}
	see Ref.~\cite{boguna2003class}, and the resulting degree distribution $P(k)$ is a mixed Poisson distribution:
	\begin{equation}
		P(k) \approx  {1 \over k!}  \int_{0}^{\R}    e^{-\langle k (\rr)\rangle }   \left[\langle k (\rr)\rangle \right]^{k} \rho_{\rr}(\rr) {\rm d} \rr 
		\label{eq:hd_pk}
	\end{equation}
	with mixing parameter $\langle k (\rr)\rangle$.
	
	The clustering coefficient of a node $i$ with degree $k_{i} > 1$ is defined as the ratio
		\begin{equation}
			c_{i} = \dfrac{t_{i}}{\binom{k_{i}}{2}}
		\end{equation}
		of the number of triangles $t_{i}$ adjacent to $i$, to the maximum such number $\binom{k_{i}}{2}$. Since nodes with degrees $k = 0$ and $k=1$ do not form triangles, their clustering coefficients are undefined. Therefore, the average clustering coefficient is defined over nodes with $k > 1$.

	The hidden variable formalism~\cite{boguna2003class} allows for expressing the average clustering coefficient of an RHG as a multiple integral over triples of hidden variables. Apart from a few special cases of the RHG, however, these integrals do not have simple closed-form solutions~\cite{candellero2016clustering}. Therefore, we restrict ourselves studying the clustering coefficient of RHGs numerically in this work.
	
	\section{Connectivity Regimes of the RHG}
	\label{sec:RHG_connectivity}
	Depending on the value of the rescaled temperature $\tau=d  T$, there exist three distinct regimes of the RHG:
	(i)~cold~($\tau < 1$), (ii)~critical~($\tau=1$), and (iii)~hot~($\tau > 1$).
	We provide detailed analyses of the properties of RHGs in these regimes below, and summarize our findings in Fig.~\ref{fig:table_full}.

	\subsection{Cold regime, $\tau < 1$} 
	\label{sec:cold}
	
	Since the inner integral in Eq.~(\ref{eq:hd_avgk2}) does not have a closed-form solution, we need to employ several approximations to derive $\langle k (\rr_{l})
	\rangle$. We note that most nodes have large radial coordinates, $e^{ \rr + \rr_{l} - \m} \gg 1$, and the dominant contribution to the inner integral in~(\ref{eq:hd_avgk2}) comes from small $\theta_1$ values.
	This allows us to estimate the integral  by replacing
	$\sin(\theta_1)$ and $\sin(\theta_1/2)$ with the leading Taylor series terms, as ${\rm sin}(x) = x + \mathcal{O}\left( x^{3} \right)$, 
	resulting in 
		\begin{align}
			&\langle k (\rr_{l}) \rangle \approx {(n-1) \pi^{d} \over
				d I_{d,1}} 
			\label{eq:hd_avgk_general}\\
			&\quad\qquad\times \int_{0}^{\R} {\rm d} \rr
			\rho_{\rr}(\rr)~_{2}F_{1}\left(1, \tau, 1+\tau, -[u_{max}(\rr_{l},\rr,\m)]^{1\over \tau}\right), \nonumber
		\end{align}
		with
		\begin{equation}
			u_{max}(\rr_{l},\rr, \m) \equiv \left({\pi \over 2}\right)^{d} e^{\rr_{l} + \rr - \m},\label{eq:hd_umax}
		\end{equation}
	and where $_{2}F_{1}$ is the Gauss' hypergeometric function, and 
	\begin{equation}
		I_{d,1} \equiv \int_{0}^{\pi} {\rm sin}^{d-1}(\theta_1) {\rm d} \theta_1 = {\sqrt{\pi} \Gamma\left[{d \over 2}\right] /\Gamma\left[{d+1 \over 2}\right],
		}
		\label{eq:id1}
	\end{equation}
	for $d > 1$, and $I_{1,1} = 2\pi$.
	
	In the $\tau < 1$ regime, the hypergeometric function in~(\ref{eq:hd_avgk_general})
	can be approximated as
		\begin{align}
			&_{2}F_{1}\left(1, \tau, 1+\tau, -[u_{max}(\rr_{l},\rr,\m)]^{\frac{1}{ \tau}}\right) \\  
			&\qquad\qquad\qquad\qquad\qquad \approx [u_{max}(\rr_{l},\rr,\m)]^{-1}{\frac{\pi \tau} {\sin\left( \pi
					\tau\right)}}, \nonumber
		\end{align}
	and $\langle k (\rr_l) \rangle$ and $\langle k \rangle$ are then approximately given by:
	\begin{align}
		\langle k \rangle &\approx  (n-1) {2^{d}\over d I_{d,1}} { \pi \tau \over \sin\left(\pi \tau\right)}
		\langle e^{-\rr} \rangle^{2} e^{\m}, 
		\label{eq:hd_cold_avg_k}\\
		\langle k (\rr) \rangle &\approx {\langle k \rangle \over \langle e^{-\rr}\rangle} e^{-\rr}, \label{eq:hd_cold_avg_kr}
	\end{align}
	where $\langle e^{-\rr}\rangle \equiv \int_{0}^{\R}{\rm d} \rr \rho_{\rr}(\rr) e^{-\rr}$, and the explicit approximation for $\langle e^{-\rr}\rangle$ follows from (\ref{eq:hd_rho_distr_rescaled}):
	\begin{equation}
		\langle e^{- \rr}\rangle \approx {\al \over \al -1} \left(e^{-\R} - e^{- \al
			\R}\right) 
		\label{eq:hd_exp_er}.
	\end{equation}
	
	We next discuss the choice of the rescaled chemical potential  $\m$. In order to do so, we discuss the leading order behavior of $\langle k\rangle_n$ in the thermodynamic limit. Since $\al > 1$, we neglect the second term in~(\ref{eq:hd_exp_er}) to obtain
	\begin{subequations}
		\begin{align}
			\langle k (\rr) \rangle_n &\sim  n e^{ \m - \rr - \R },
			\label{eq:hd_cold_avg_k_scaling}\\
			\langle k\rangle_n & \sim  n e^{ \m - 2\R }. 
			\label{eq:hd_cold_avg_kr_scaling}
		\end{align}
	\end{subequations}

	Henceforth, we write $f(x) \sim g(x)$ when $\lim_{x \to \infty} \frac{f(x)}{g(x)} = K$, where $K>0$ is a constant.
	
	We note that $\langle k (\rr) \rangle$ decreases exponentially as a function of $\rr$ with the largest (smallest) expected
	degree corresponding to $\rr=0$ ($\rr=\R$). By demanding
	that the largest and smallest expected degrees scale as
	\begin{subequations}
		\begin{align}
			\langle k_{max} \rangle_n &= \langle k(0)  \rangle_n \sim  n,
			\label{eq:scaling_assumption1}\\
			\langle k_{min} \rangle_n &= \langle k\left(\R\right)  \rangle_n \sim  1,
			\label{eq:scaling_assumption2}
		\end{align}
	\end{subequations}
	we obtain $\R \sim \ln n$ and $\m = \R + \lambda$, where $\lambda$ is an arbitrary constant.
	
	First, we note that the scaling for $\R$ is consistent with our initial assumption of $\R \gg 1$
	for large graphs. We also note that the exact value of the  parameter $\lambda$
	is not important as long as it is independent of $n$. To be consistent with the original $\mathbb{H}^{2}$ formulation we set $\lambda =0$, obtaining
	\begin{equation}
		\m = \R = \ln\left(n/\nu\right),
		\label{eq:hd_cold_scaling}
	\end{equation}
	where $\nu$ is a parameter controlling the expected degree of the RHG.
	
	Applying the scaling relationships (\ref{eq:hd_cold_scaling})  to (\ref{eq:hd_cold_avg_k}) and (\ref{eq:hd_cold_avg_kr}),  
	we obtain
	\begin{align}
		\langle k \rangle_n & \approx \nu  {2^{d} \over d I_{d,1}} \left({\al \over \al - 1}\right)^{2} {\pi \tau \over \sin(\pi \tau)} \nonumber\\
		&\qquad \times \left[1 - 2 \left(\frac{n}{\nu}\right)^{1-\al} + \left(\frac{n}{\nu}\right)^{2(1-\al)}\right], 
		\label{eq:hd_cold_avg_k2}
	\end{align}
	and 
	\begin{equation}
		\label{eq:hd_cold_avg_kr2}
		\langle k(\rr) \rangle_n \approx {n \over \nu} {\al -1 \over \al}  \langle k \rangle \,e^{-\rr} \sum_{\ell\,=\,0}^{\infty}  \left(\frac{\nu}{n}\right)^{\ell (\al-1)},
	\end{equation}
	see Fig.~\ref{fig:kr_plots}(a-c).
	
\begin{figure}
		\includegraphics[width=3in]{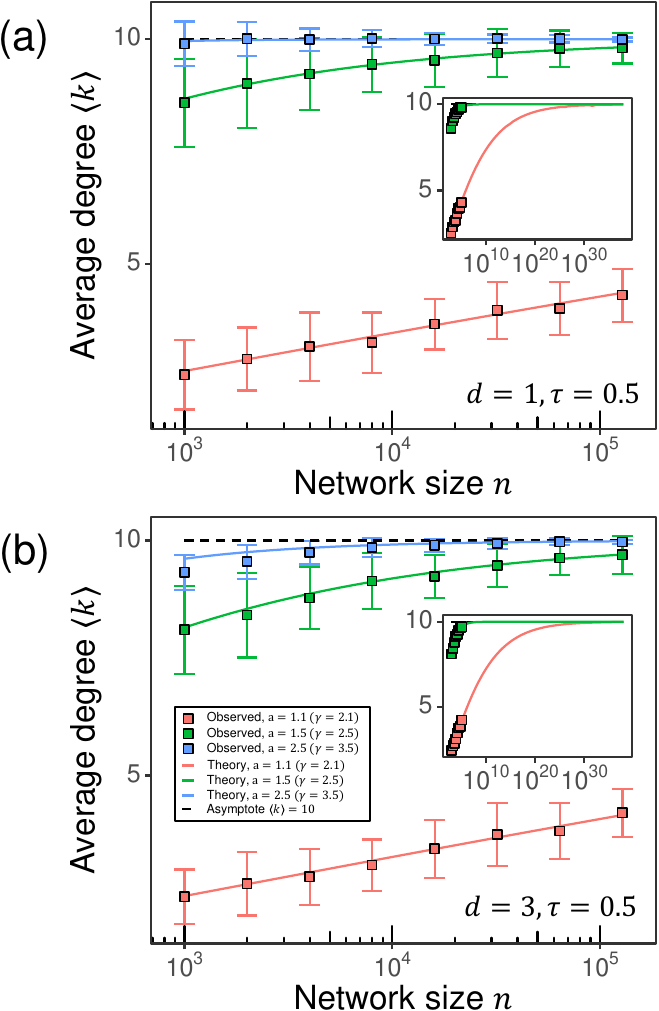}
		\caption{ \footnotesize  Expected degree $\langle k \rangle$ as a function of network size $n$ for RHGs in the cold ($\tau = 0.5$) regime with (a) $d=1$ and (b) $d=3$. Each panel includes the results for (red) $\al=1.1$ ($\gamma = 2.1$),  (green) $\al=1.5$ ($\gamma = 2.5$), and (blue)  $\al=2.5$ ($\gamma = 3.5$). Each point is the average of $100$ simulations and the error bars display standard deviations. Solid lines are theoretical values for $\langle k \rangle$ prescribed by Eqs.~(\ref{eq:hd_avgk2}) and~(\ref{eq:hd_avgk3}) and the dashed line is the thermodynamic limit of Eq.~(\ref{eq:hd_cold_avg_k2}). The insets in (a) and (b) correspond to extended domains of $n$ values for the cases $\al=1.1$ and $\al=1.5$. Note that the $\al=1.1$ case converges to the asymptotic value at a much slower rate  compared to the  $\al=1.5$ and $\al=2.5$ cases.
		} 
		\label{fig:avgkcold}
	\end{figure}

	As seen from Eq.~(\ref{eq:hd_cold_avg_k2}) and Fig.~\ref{fig:avgkcold}, RHGs are sparse in the cold~($\tau<1$) regime. Henceforth, we call graphs sparse if their expected  degree converges to a finite constant in the thermodynamic limit. The slow convergence to the asymptotic value of $\langle k \rangle=10$ in the $\al=1.1$ case is due to the breakdown of the ${\rm sin}(x) \approx x$ approximation in Eq.~(\ref{eq:hd_avgk_general}) for small $\al$ values. Indeed, at small $\al$ values a larger fraction of nodes is characterized by small $\rr$ values, for which  the $e^{ \left( \rr + \rr_{l} - \m \right)} \gg 1$ assumption fails.
	
\begin{figure}
		\includegraphics[width=3in]{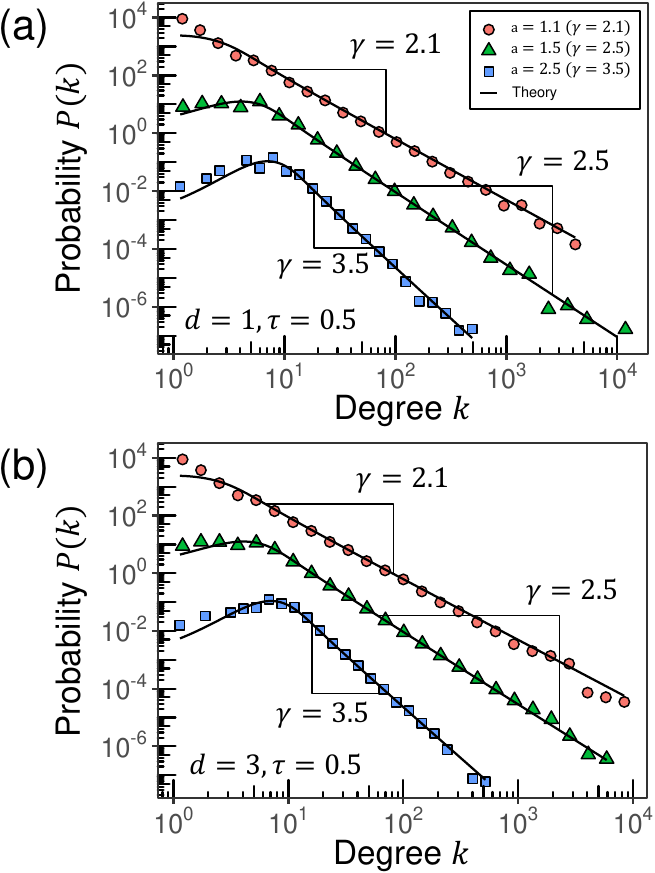}
		\caption{\footnotesize Degree distribution $P(k)$ for RHGs with (a) $d=1$ and (b) $d=3$ at $\tau = 0.5$ and $n = 1000 \cdot 2^{7}$. Each panel includes the degree distributions for (red) $\al=1.1$ ($\gamma = 2.1$),  (green) $\al=1.5$ ($\gamma = 2.5$), and (blue) $\al=2.5$ ($\gamma = 3.5$). Probabilities $P(k)$ are obtained from a single network realization. Degree distributions are binned logarithmically to suppress noise at large $k$ values. Solid lines are theoretical values for $P(k)$ prescribed by Eq.~(\ref{eq:pk_plaw}). For visibility, the probabilities corresponding to $\gamma = 2.5$ and $\gamma = 2.1$ are multiplied by $10^{2}$ and $10^{4}$, respectively. The scaling constant $\nu = 10 \times {d I_{d,1} \over 2^{d}} \left({\al - 1\over \al}\right)^{2} {\sin(\pi \tau) \over \pi \tau}$, corresponding to $\langle k \rangle$ = 10 in the thermodynamic limit.
		} 
		\label{fig:pk_cold}
	\end{figure}

	Finally, using Eqs.~(\ref{eq:hd_cond_pk}) and (\ref{eq:hd_pk}) we obtain the Pareto-mixed Poisson distribution, which is a power law
		\begin{equation}
			P(k) \approx \al \kappa_{0}^{\al} \, {\Gamma[k - \al, \kappa_{0}] \over \Gamma[k + 1]} \sim k^{-(\al+1)},
			\label{eq:pk_plaw}
		\end{equation}
		where $\Gamma[s,x]$ is the upper incomplete gamma function, and $\kappa_{0} \equiv  \left(\frac{ \al-1 } {\al }\right) \langle k \rangle$, as confirmed by simulations in Fig.~\ref{fig:pk_cold}.
	
	Hence, the cold regime corresponds to sparse power-law graphs, $P(k) \sim k^{-\gamma}$, with $\gamma = \al + 1 \in (2,\infty)$. We note that the degree distribution is a power law if it takes the form of $P(k) = \ell(k) k^{-\gamma}$, where $\ell(k)$ is a slowly varying function, i.e., a function that varies slowly at infinity, see Ref.~\cite{voitalov2019scale}. Any function converging to a constant is slowly varying, and, in the case of Eq.~(\ref{eq:pk_plaw}), $\ell(k) \to \al \kappa_{0}^{\al} $ as $k \to \infty$. The degree distribution is called scale-free if $\gamma \in (2,3)$.
	
	The special case of $\al = 1$ ($\gamma = 2$) is also well defined. In this  case  the expressions for $\langle k \rangle$ and $\langle k (\rr)\rangle$ given by Eqs.~(\ref{eq:hd_cold_avg_k})~and~(\ref{eq:hd_cold_avg_kr}) remain valid but $\langle e^{-\rr}\rangle$ is now given by
	\begin{equation}
		\langle e^{- \rr} \rangle \approx \R e^{-\R}.
		\label{eq:g2expmr}
	\end{equation}

\begin{figure}
		\includegraphics[width=3in]{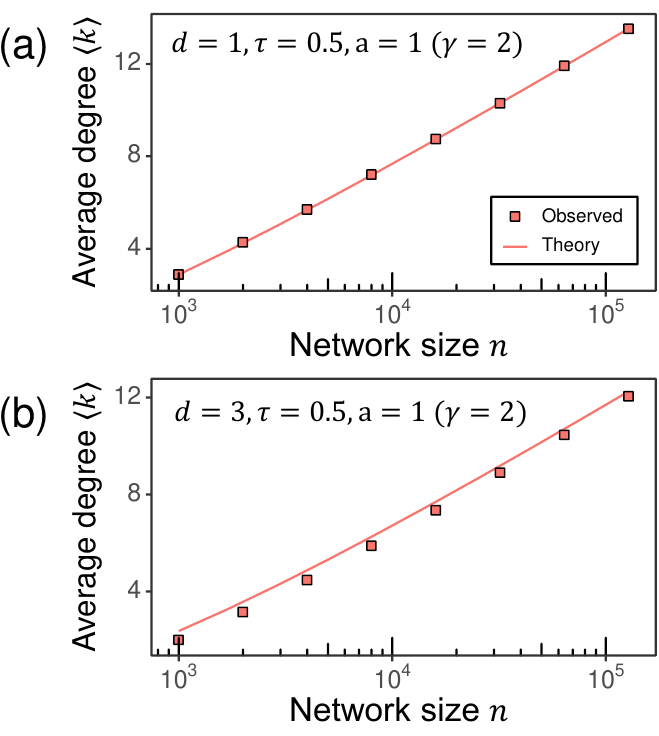}
		\caption{\footnotesize Expected degree $\langle k \rangle$ of RHGs with (a) $d=1$ and (b) $d=3$ in the case of $\al = 1$ at $\tau = 0.5$.  Each point is the average of $100$ simulations and the error bars display standard deviations (not visible for most points here). To avoid fluctuations associated with large degree nodes, we have imposed a cutoff in the radial coordinate distribution, removing nodes with $\rr \leq \rr_{\textrm{cut}}$, where $\rr_{\textrm{cut}}$ is defined such that $\langle k(\rr_{\textrm{cut}}) \rangle = n^{1/2}$.  Solid lines are theoretical values for $\langle k \rangle$ prescribed by Eq.~(\ref{eq:hd_cold_gamma2_avg_k2}), corrected for the radial coordinate cutoff. The scaling constant is set to $\nu = 10 \times {d I_{d,1} \over 2^{d}} {\sin(\pi \tau) \over \pi \tau}$. 
		}
		\label{fig:avgk_cold_a1}
	\end{figure}

\begin{figure}
		\includegraphics[width=3in]{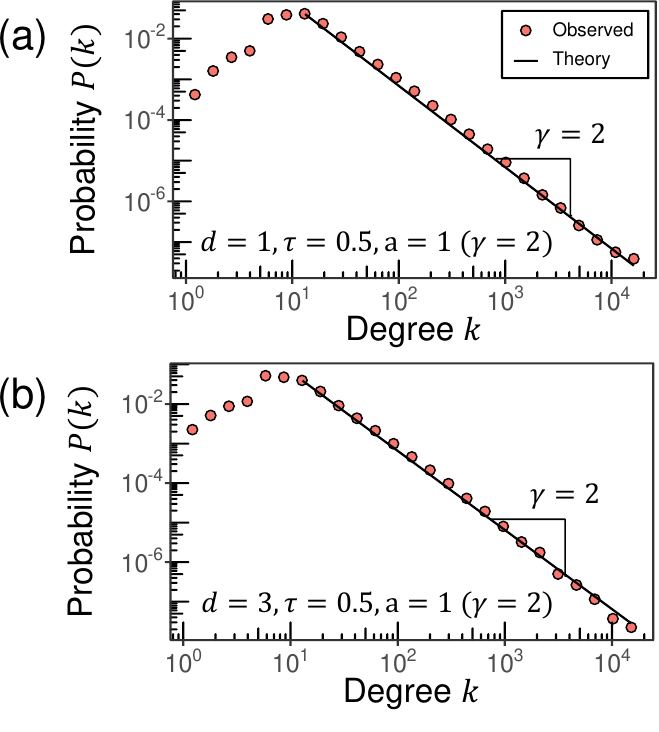}
		\caption{\footnotesize Degree distribution $P(k)$ for RHGs with (a) $d=1$ and (b) $d=3$ in the case of $\al = 1$ at $\tau = 0.5$ and $n = 1000 \cdot 2^{7}$. Probabilities $P(k)$ are obtained from a single network realization. Degree distributions are binned logarithmically to suppress noise at large $k$ values. To avoid fluctuations associated with large degree nodes, we have imposed a cutoff in the radial coordinate distribution, removing nodes with $\rr \leq \rr_{\textrm{cut}}$, where $\rr_{\textrm{cut}}$ is defined such that $\langle k(\rr_{\textrm{cut}}) \rangle = n^{1/2}$. Solid lines are theoretical values for $P(k)$ based on a slope of $\gamma = 2$. The scaling constant is set to $\nu = 10 \times {d I_{d,1} \over 2^{d}} {\sin(\pi \tau) \over \pi \tau}$.
		}
		\label{fig:degdist_cold_a1}
	\end{figure}

	It is straightforward to verify that the scaling of $\R = \m = \ln \left (n/\nu \right)$ in the $\al =1$ case does not lead to the desired calibration of node degrees, $\langle k_{max} \rangle \sim n$ and $\langle k_{min} \rangle \sim  1$. Instead, the proper scaling is
	\begin{subequations}
		\begin{align}
			\R &= {\rm ln} \left( n/ \nu\right),
			\label{eq:scaling_cold_g2_r}\\
			\m &= \R-   \rm \ln  \R,
			\label{eq:scaling_cold_g2_m}
		\end{align}
	\end{subequations}
	resulting in 
	\begin{eqnarray}
		\label{eq:hd_cold_gamma2_avg_k2}
		\langle k \rangle_{n} & \approx & \nu  {2^{d} \over d I_{d,1}} {\pi \tau \over \sin(\pi \tau)} {\rm ln} \left(n / \nu\right),\\
		\langle k(\rr) \rangle_{n} & \approx & {n \over \nu} \frac{\langle k \rangle}{ {\rm ln} \left(n / \nu\right)} e^{-\rr}.
	\end{eqnarray}
	In other words, the $\al=1$ ($\gamma = 2$) case corresponds to graphs with $\langle k \rangle_{n} \sim \ln\left(n / \nu\right)$, as confirmed by Fig.~\ref{fig:avgk_cold_a1}. Degree distribution $P(k)$ matches a power law with $\gamma = 2$, as shown in Fig.~\ref{fig:degdist_cold_a1}. The phenomenon that RHGs are no longer sparse in the $\al=1$ ($\gamma =2$) case is not specific to the model but is a general property of all scale-free network models with $P(k) \sim k^{-2}$.
	
\begin{figure}
		\includegraphics[width=3in]{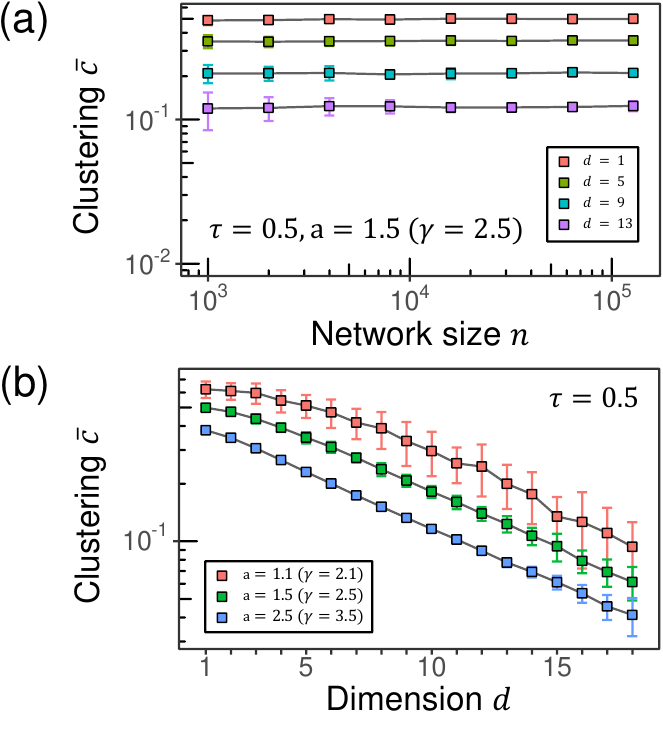}
		\caption{ \footnotesize Average clustering coefficient $\bar{c}$ of nodes with degree $k > 1$ as a function of network size (a) and as a function of dimension $d$ (b) at $\tau = 0.5$. Panel (a) includes results for (red) $d = 1$, (green) $d = 5$, (blue) $d = 9$ and (purple) $d = 13$, for $\al = 1.5$ ($\gamma = 2.5$), while panel (b) includes results for (red) $\al=1.1$ ($\gamma = 2.1$),  (green) $\al=1.5$ ($\gamma = 2.5$), and (blue) $\al=2.5$ ($\gamma = 3.5$), for $n = 10^{4}$. Each point is the average of $100$ simulations and the error bars display standard deviations. The scaling constant $\nu = 10 \times {d I_{d,1} \over 2^{d}} \left({\al - 1\over \al}\right)^{2} {\sin(\pi \tau) \over \pi \tau}$, corresponding to $\langle k \rangle$ = 10 in the thermodynamic limit.}
		\label{fig:cold_clust}
	\end{figure}

	Here, we do not attempt to obtain analytical expressions of clustering for $d \geq 1$, as its computation is quite involved already in the $d=1$ case \cite{candellero2016clustering, vanderkolk2022anomalous}, but we inspect average clustering numerically as a function of dimension and as a function of network size in Fig.~\ref{fig:cold_clust}. In the cold regime $\tau < 1$, we observe that clustering is nonvanishing in the thermodynamic limit, similar to the $d=1$ case. The average clustering decreases as a function of dimension $d$. This property of the clustering coefficient has been observed in other spatial graph models as well, particularly, GIRGS~\cite{almagro2022detecting}. This is because at $d \rightarrow \infty$, the angular distance distribution between two random points on the unit $d$-sphere~\cite{hammersley1950} approaches the Dirac delta function centered at~$\pi/2$. As a result, the role of a node's angular coordinates in the hyperbolic distances diminishes, and the network becomes ``less geometric" and more similar to the hyper soft configurational model~(HSCM), Sec.~\ref{sec:dinf}.

		\subsubsection*{Graph property perspective, $\tau < 1$}
		From a graph property viewpoint, the RHG is instrumental in generating synthetic networks with desired properties. It follows from our analysis that RHGs in the cold regime $\tau < 1$ are characterized by power-law degree distributions, $P(k) \sim k^{-\gamma}$, where the exponent~$\gamma \in [2, \infty)$ is a function of RHG temperature~$\tau$ and node density parameter~$\al$. The radius $\R$ of the hyperbolic ball~$\mathbb{B}^{d+1}$, on the other hand, controls the expected degree~$\langle k \rangle$ and the sparsity of the resulting graphs. Relying on these results, we can redefine the RHG model in terms of the parameters $\left(n, \gamma, \tau, \langle k \rangle \right)$. 
		
		To generate an RHG in the cold regime with desired graph properties for $\gamma > 2$, one sets the node density parameter $\al$, the chemical potential $\m$ and radius $\R$ of~$\mathbb{B}^{d+1}$ to
		\begin{subequations}
			\begin{align}
				\al &= \gamma - 1,\\
				\m &= \R = \ln \left(n / \nu \right),
			\end{align}
		\end{subequations}
		where $\nu$ is the solution of Eq.~(\ref{eq:hd_cold_avg_k2}), now taking the form of 
		\begin{align}
			\langle k \rangle & \approx \nu  {2^{d} \over d I_{d,1}} \left({\gamma - 1 \over \gamma - 2}\right)^{2} {\pi \tau \over \sin(\pi \tau)} \nonumber\\
			&\qquad \times \left[1 - 2 \left(\frac{n}{\nu}\right)^{2-\gamma} + \left(\frac{n}{\nu}\right)^{2(2-\gamma)}\right].
			\label{eq:hd_cold_avg_k2_user}
		\end{align}
		When $\gamma = 2$ in the cold regime, one must set
		\begin{subequations}
			\begin{align}
				\al &= \gamma - 1 = 1,\\
				\R &= \ln \left(n / \nu \right),\\
				\m &= \R- \ln \R,
			\end{align}
		\end{subequations}
		and $\nu$ is obtained as the solution of Eq.~(\ref{eq:hd_cold_gamma2_avg_k2}), which now takes the form of
		\begin{equation}
			\langle k \rangle \approx \nu  {2^{d} \over d I_{d,1}} {\pi \tau \over \sin(\pi \tau)} {\rm ln} \left(n / \nu\right). \label{eq:hd_cold_g2_avg_k2_user}
		\end{equation}

	\subsection{Critical regime, $\tau=1$}
	\label{sec:critical}
	In the $\tau=1$ regime, Eqs.~(\ref{eq:hd_avgk_general}) and (\ref{eq:hd_avgk3}) can be approximated
	as:
	\begin{align}
		\langle k\rangle_n & \approx   { (n-1) 2^{d} \over d I_{d,1}  }  e^{\m} \label{eq:hd_crit_avg_k}\\
		&\times \left[d\ln \left( {\pi \over 2} \right) \langle e^{-\rr} \rangle^{2}  + 2 \langle e^{-\rr} \rangle \langle \rr e^{-\rr} \rangle - \m  \langle e^{-\rr} \rangle^{2} \right],\nonumber\\
		\langle k(\rr)\rangle_n & \approx  { (n-1) 2^{d} \over d I_{d,1}  }  e^{\m - \rr}
		\label{eq:hd_crit_avg_kr}\\
		&\times \left[d\ln \left( {\pi \over 2} \right) \langle e^{-\rr} \rangle  + (\rr - \m) \langle e^{-\rr} \rangle + \langle \rr e^{-\rr} \rangle \right],\nonumber
	\end{align}
	where  $\langle \rr e^{-\rr} \rangle \equiv \int {\rm d} \rr \rho_{\rr}(\rr) \rr e^{-\rr}$ is given by
	\begin{equation}
		\langle \rr e^{-\rr} \rangle \approx \left(\dfrac{\al}{\al-1}\right) \left[\left(\R - \dfrac{1}{\al-1}\right) e^{-\R} + \dfrac{1}{\al-1} e^{-\al\R}\right].
		\label{eq:hd_exp_rer}
	\end{equation}
	Given that $\al > 1$, we drop the second terms in (\ref{eq:hd_exp_er}) and (\ref{eq:hd_exp_rer}) to obtain:
	\begin{align}
		\langle k(\rr)\rangle_n & \approx  { (n-1) 2^{d} \over d I_{d,1}  } \left(\dfrac{\al}{\al-1}\right)
		\label{eq:hd_crit_avg_kr2}\\
		&\times \left(d \log \left(\frac{\pi}{2}\right) - {1 \over \al - 1} + \R - \m  + \rr\right)  e^{\m-\R-\rr}.\nonumber
	\end{align}
	Similar to the $\tau<1$ regime, we demand $\langle k_{max} \rangle_n \sim n$ and $\langle k_{min} \rangle_n \sim 1$ to
	obtain the scaling relationships for $\m$ and $\R$. For $\langle k_{max} \rangle_n = \langle k(0) \rangle_n$ and $\langle k_{min} \rangle_n = \langle k(\R) \rangle_n$, we have
	\begin{subequations}
		\begin{align}
			\langle k(0) \rangle_n &\sim  n \left(d \log \left(\frac{\pi}{2}\right) - {1 \over \al - 1} + \R - \m\right) e^{ \m - \R },
			\label{eq:hd_crit_avg_k_scaling1}\\
			\langle k(\R) \rangle_n & \sim  n \left(d \log \left(\frac{\pi}{2}\right) - {1 \over \al - 1} + 2\R - \m\right) e^{ \m - 2\R }.
			\label{eq:hd_crit_avg_k_scaling2}
		\end{align}
	\end{subequations}
	Scaling $\langle k (0) \rangle_n \sim  n$ and  $\langle k (\R) \rangle_n \sim  1$ is achieved when $\m = \R$ and $\R e^{-\R} \sim {1 \over n}$. Analogous to the cold regime, we set $\R^{-1} e^{\R} = {n \over \nu}$, obtaining
	\begin{equation}
		\m = \R = -W_{-1}\left(-\frac{\nu}{n}\right), \label{eq:hd_critical}
	\end{equation}
	where $W_{-1}(\cdot)$ is the $W_{-1}$ branch of the Lambert $W$ function.
	
\begin{figure}
		\includegraphics[width=3in]{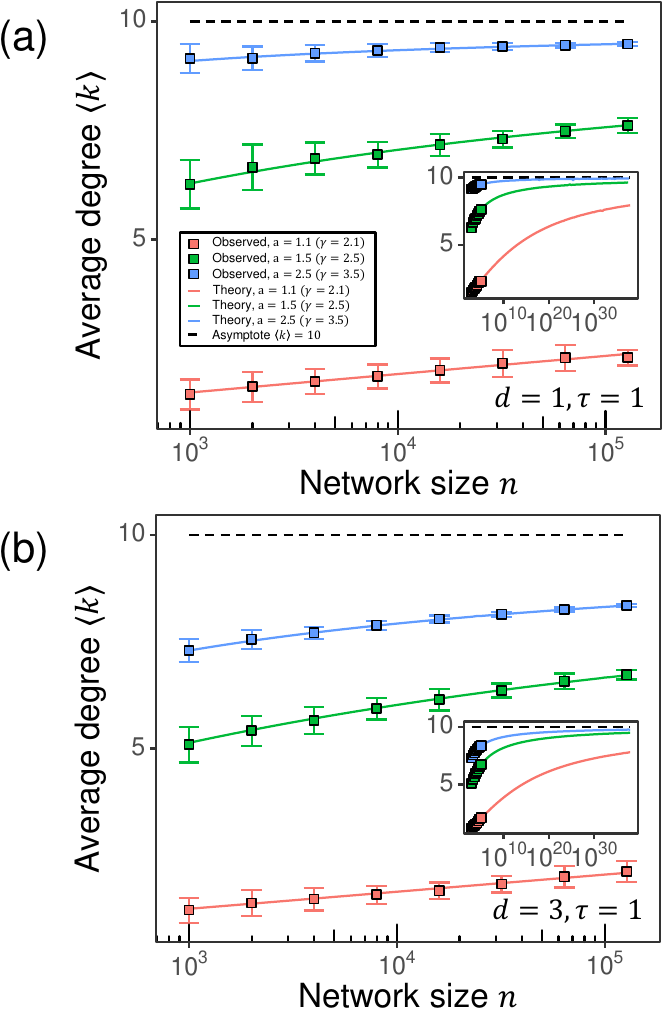}
		\caption{ \footnotesize Expected degree $\langle k \rangle$ as a function of network size $n$ for RHGs in the critical ($\tau = 1$) regime with (a) $d=1$ and (b) $d=3$. Each panel includes the results for (red) $\al=1.1$ ($\gamma = 2.1$),  (green) $\al=1.5$ ($\gamma = 2.5$), and (blue)  $\al=2.5$ ($\gamma = 3.5$). Each point is the average of $100$ simulations and the error bars display standard deviations. Solid lines are theoretical values for $\langle k \rangle$ prescribed by Eqs.~(\ref{eq:hd_avgk2}) and~(\ref{eq:hd_avgk3}) and the dashed line is the thermodynamic limit of Eq.~(\ref{eq:hd_crit_avg_k2}). The insets in (a) and (b) correspond to extended domains of $n$ values. Note that in the critical regime all 3 cases converge to the asymptotic value at a much slower rate than in the cold regime, while the $\al = 1.1$ case has not yet converged even within the extended domain of $n$ values.}
		\label{fig:avgk_critical}
	\end{figure}

	Using the scaling in (\ref{eq:hd_critical}), we obtain
	\begin{align}
		\langle k \rangle_n & \approx \dfrac{2^{d}}{d I_{d,1}} \left({\al \over \al - 1}\right)^{2} 
		\label{eq:hd_crit_avg_k2} \\
		&\times \left[ 1 - \left(d \log \left(\frac{\pi}{2}\right) - \dfrac{2}{\al-1}\right) \left(W_{-1}\left(-{\nu \over n}\right)\right)^{-1}\right],\nonumber\\
		\langle k(\rr)\rangle_n & \approx  n { 2^{d} \over d I_{d,1}  }
		\left({\al \over \al - 1}\right) \left(d \log \left(\frac{\pi}{2}\right) - {1 \over \al - 1}  + \rr\right) e^{- \rr},
		\label{eq:hd_crit_avg_kr2}
	\end{align}
	see Fig.~\ref{fig:kr_plots}(d-f).
	Hence, the critical regime corresponds to sparse graphs in the thermodynamic limit, as confirmed by simulations in Fig.~\ref{fig:avgk_critical}. Note that the convergence to the asymptote of $\langle k \rangle = 10$ is slower than in the cold regime, likely due to relatively large subleading terms in Eq.~(\ref{eq:hd_crit_avg_k2}).
	
\begin{figure}
		\includegraphics[width=3in]{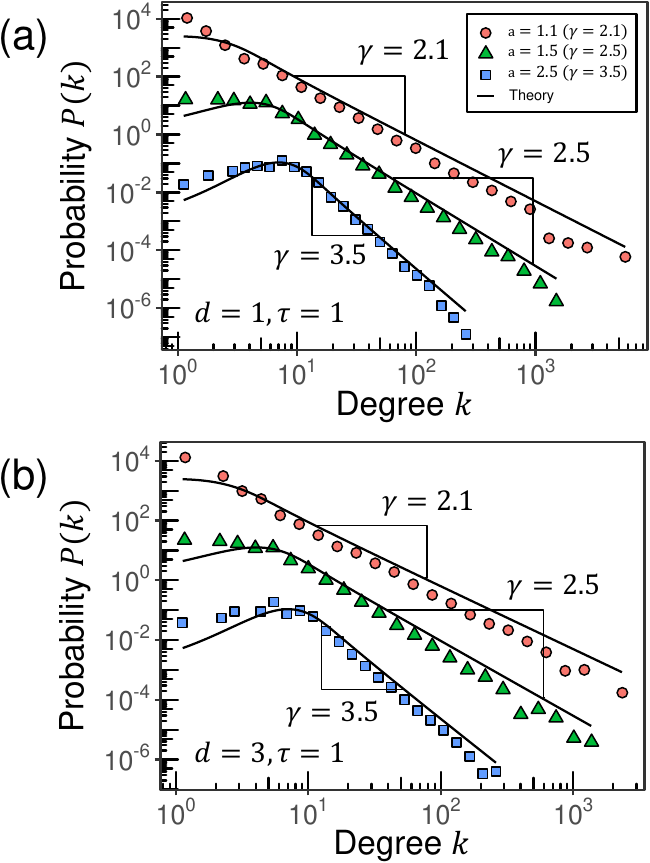}
		\caption{\footnotesize Degree distribution $P(k)$ for RHGs with (a) $d=1$ and (b) $d=3$ at $\tau = 1$ and $n = 1000 \cdot 2^{7}$. Each panel includes the degree distributions for (red) $\al=1.1$ ($\gamma = 2.1$),  (green) $\al=1.5$ ($\gamma = 2.5$), and (blue)  $\al=2.5$ ($\gamma = 3.5$). Probabilities $P(k)$ are obtained from a single network realization. Degree distributions are binned logarithmically to suppress noise at large $k$ values. For visibility, the probabilities corresponding to $\gamma = 2.5$ and $\gamma = 2.1$ are multiplied by $10^{2}$ and $10^{4}$, respectively. Solid lines are theoretical values for $P(k)$ prescribed by Eq.~(\ref{eq:pk_plaw}). The scaling constant $\nu$ is chosen such that $ \nu = 10 \times {d I_{d,1} \over 2^{d}} \left({\al - 1\over \al}\right)^{2}$, corresponding to $\langle k \rangle$~=~10 in the thermodynamic limit.}
		\label{fig:pk_critical}
	\end{figure}

	We observe that the degree distributions of RHGs in the critical regime seem to follow a power law with the same exponent as in the cold regime:
	\begin{equation}
		\begin{aligned}
			P(k) &\sim k^{-\gamma},\\
			\gamma & = \al + 1,
		\end{aligned}
	\end{equation}
	as observed from Fig.~\ref{fig:pk_critical}. This is the case since the tail of $P(k)$ is dominated by nodes with small $\rr$ values. In the critical regime, $\langle k(\rr)\rangle_n \sim e^{-\rr}$ for $\rr$ values close to $0$, similar to the 
	cold regime, resulting in the same degree distribution exponent $\gamma$.
	
\begin{figure}
		\includegraphics[width=3in]{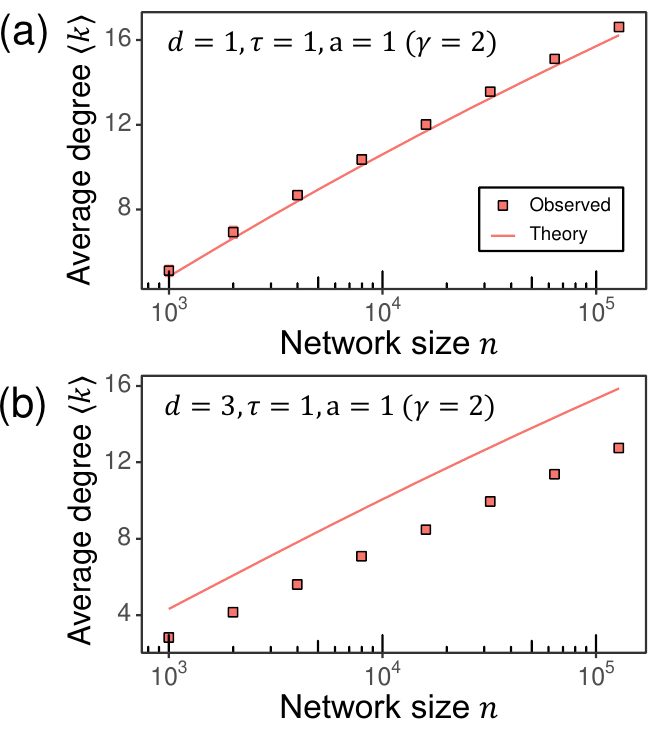}
		\caption{\footnotesize Expected degree $\langle k \rangle$ of RHGs with (a) $d=1$ and (b) $d=3$ in the case of $\al = 1$ at $\tau = 1$.  Each point is the average of $100$ simulations and the error bars display standard deviations (not visible for most points here). To avoid fluctuations associated with large degree nodes, we have imposed a cutoff in the radial coordinate distribution, removing nodes with $\rr \leq \rr_{\textrm{cut}}$, where $\rr_{\textrm{cut}}$ is defined like in the cold regime at $\tau = 0.5$, as Eq.~(\ref{eq:hd_crit_gamma2_avg_kr}) cannot be solved exactly. Solid lines are theoretical values for $\langle k \rangle$ prescribed by Eq.~(\ref{eq:hd_crit_gamma2_avg_k}), corrected for the radial coordinate cutoff. The scaling constant is set to $\nu = 10 \times {d I_{d,1} \over 2^{d}}$.}
		\label{eq:fig_avgk_crit_a1}
	\end{figure}

\begin{figure}
		\includegraphics[width=3in]{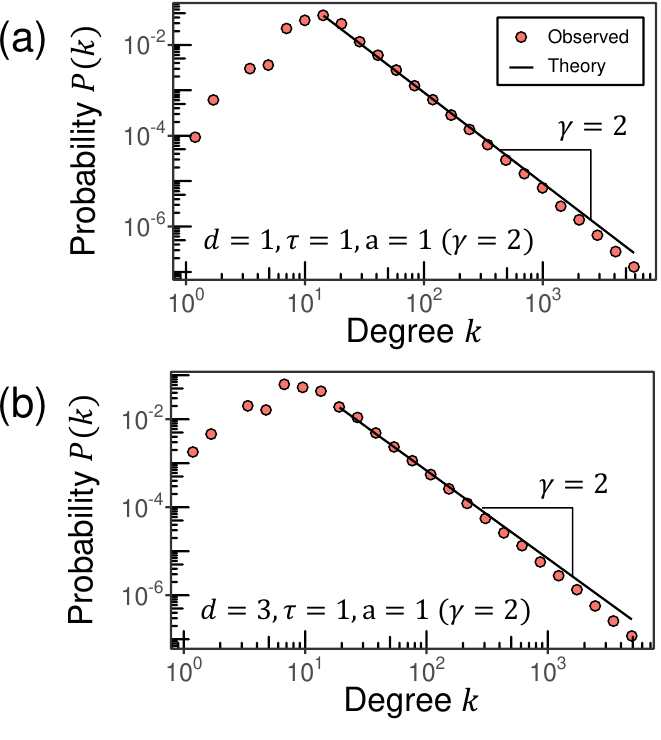}
		\caption{\footnotesize Degree distribution $P(k)$ for RHGs with (a) $d=1$ and (b) $d=3$ in the case of $\al = 1$ at $\tau = 1$ and $n = 1000 \cdot 2^{7}$. Probabilities $P(k)$ are obtained from a single network realization. Degree distributions are binned logarithmically to suppress noise at large $k$ values. To avoid fluctuations associated with large degree nodes, we have imposed a cutoff in the radial coordinate distribution, removing nodes with $\rr \leq \rr_{\textrm{cut}}$, where $\rr_{\textrm{cut}}$ is defined like in the cold regime at $\tau = 0.5$, as Eq.~(\ref{eq:hd_crit_gamma2_avg_kr}) cannot be solved exactly. Solid lines are theoretical values for $P(k)$ based on a slope of $\gamma = 2$. The scaling constant is set to $\nu = 10 \times {d I_{d,1} \over 2^{d}}$.
		}
		\label{fig:degdist_crit_a1}
	\end{figure}

	To investigate the $\al =1$ ($\gamma=2$) case of the critical regime, we need to reexamine the scaling of $\langle k(\rr = 0) \rangle_{n}$ and $\langle k(\rr = \R) \rangle_{n}$.
	To do so, we use Eq.~(\ref{eq:hd_avgk2}) with $\tau=1$ and $\al =1$, arriving, to the leading order, at
	\begin{subequations}
		\label{eq:scaling_crit_gamma2}
		\begin{align}
			\label{eq:scaling_crit_gamma2_k0}
			\langle k(\rr = \R)\rangle & \approx \frac{2^{d} \left(\frac{\pi}{2}\right)^{d}}{I_{d,1}}n,\\
			\label{eq:scaling_crit_gamma2_kr}
			\langle k(\rr = 0)\rangle & \approx \frac{3 \times 2^{d-1} \left(\frac{\pi}{2}\right)^{d}}{I_{d,1}}n \R^{2} e^{\m - 2\R}.
		\end{align}
	\end{subequations}
	It is seen from Eq.~(\ref{eq:scaling_crit_gamma2}) that the desired scalings of $\langle k_{max} \rangle_n = \langle k(\rr = 0) \rangle_n \sim n$ and $\langle k_{min} \rangle_n = \langle k(\rr =\R) \rangle_n ~ \sim 1$ are obtained if we set 
	$\R = \ln \left(n/\nu\right)$, and $\m = \R - 2 \ln \R$. Then,  

	\begin{align}
		\langle k(\rr) \rangle_n  & \approx   { 2^{d} \over d I_{d,1}  } \frac{n}{\left[\ln \left (n / \nu \right)\right]^{2}}  e^{- \rr} \nonumber\\
		&\quad\times  \left[   {\rm Li}_{2}\left[-  \left[\ln \left (\frac{n} { \nu} \right)\right]^{2} \left(\frac{\pi}{2}\right)^{d} e^{\rr-\R} \right]\right.   \nonumber\\
		&\qquad\qquad\left. - \, {\rm Li}_{2}\left[- \left[\ln \left (\frac{n} { \nu} \right)\right]^{2}  \left(\frac{\pi}{2}\right)^{d} e^{\rr} \right] \right],
		\label{eq:hd_crit_gamma2_avg_kr}
	\end{align}
	\begin{align}
		\langle k \rangle_n  & \approx \nu { 2^{d} \over d I_{d,1}  } \frac{1}{\left[\ln\left(n / \nu \right)\right]^{2}}  \nonumber\\
		&\quad\times  \left[   2{\rm Li}_{3}\left[-  \left[\ln \left (\frac{n} { \nu} \right)\right]^{2} \left(\frac{\pi}{2}\right)^{d} \right]\right. \nonumber\\
		&\qquad\qquad - {\rm Li}_{3}\left[- \left[\ln \left (\frac{n} { \nu} \right)\right]^{2}  \left(\frac{\pi}{2}\right)^{d} e^{-\R} \right] \nonumber\\
		&\qquad\qquad\left. - \, {\rm Li}_{3}\left[- \left[\ln \left (\frac{n} { \nu} \right)\right]^{2}  \left(\frac{\pi}{2}\right)^{d} e^{\R} \right] \right],
		\label{eq:hd_crit_gamma2_avg_k}
	\end{align}
	and
	\begin{equation}
		\langle k \rangle_n \sim \nu \frac{ 2^{d}}{d I_{d,1}} \ln \left(\frac{n}{\nu}\right),
	\end{equation}
	where ${\rm Li}_{s}(x)$ is the $s$-th order polylogarithm function. Like in the cold regime, the $\al = 1$ case in the critical regime corresponds to graphs with $\langle k \rangle_{n} \sim \ln\left(n / \nu\right)$, as confirmed by Fig.~\ref{eq:fig_avgk_crit_a1}. We note that there is a disagreement of theoretical and simulated values for $d>1$ in Fig.~\ref{eq:fig_avgk_crit_a1}(b), likely caused by breakdown of the approximation of $[\sin(\theta_{1}/2)]^{d}$ in Eq.~(\ref{eq:hd_avgk2}) and imperfect control of $\R$ and $\m$ at $\tau=1, \al=1$. The degree distribution for $\al = 1$ in the critical regime is shown in Fig.~\ref{fig:degdist_crit_a1}.
	
\begin{figure}
		\includegraphics[width=3in]{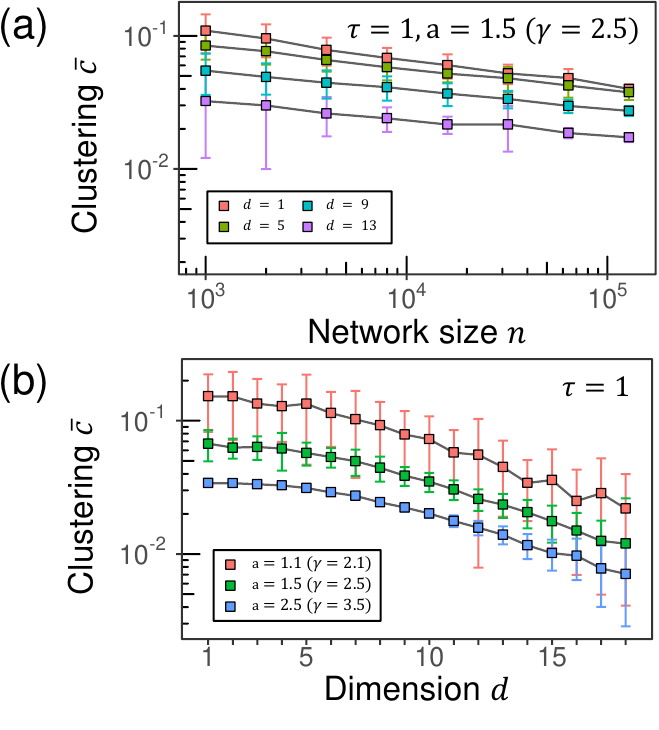}
		\caption{ \footnotesize{ Average clustering coefficient $\bar{c}$ (a) of nodes with degree $k > 1$ as a function of network size $n$ and (b) as a function of dimension $d$  at $\tau = 1$. Panel (a) includes results for (red) $d = 1$, (green) $d = 5$, (blue) $d = 9$ and (purple) $d = 13$, for $\al = 1.5$ ($\gamma = 2.5$), while panel (b) includes results for (red) $\al=1.1$ ($\gamma = 2.1$),  (green) $\al=1.5$ ($\gamma = 2.5$), and (blue) $\al=2.5$ ($\gamma = 3.5$), for $n = 10^{4}$. Each point is the average of $100$ simulations and the error bars display standard deviations. The scaling constant $\nu$ is chosen such that $ \nu = 10 \times {d I_{d,1} \over 2^{d}} \left({\al - 1\over \al}\right)^{2}$, corresponding to $\langle k \rangle$~=~10 in the thermodynamic limit.}}
		\label{fig:critical_clust}
	\end{figure}

	Unlike the cold regime, Fig.~\ref{fig:critical_clust}(a) shows that average clustering in the critical regime decreases logarithmically with $n$ as in the $d=1$ case~\cite{vanderkolk2022anomalous}. Similar to the cold regime, we observe that the average clustering coefficient decreases as dimensionality $d$ increases, Fig.~\ref{fig:critical_clust}(b).

		\subsubsection*{Graph property perspective, $\tau = 1$}
		To generate an RHG in the critical regime with desired graph properties for $\gamma > 2$, one must set
		\begin{subequations}
			\begin{align}
				\al &= \gamma -1,\\
				\m &= \R = -W_{-1}\left(-{\nu \over n}\right),
			\end{align}
		\end{subequations}
		where $\nu$ is determined by Eq.~(\ref{eq:hd_crit_avg_k2}), which now takes the form of
		\begin{align}
			\langle k \rangle & \approx \dfrac{2^{d}}{d I_{d,1}} \left({\gamma-1 \over \gamma - 2}\right)^{2}  \label{eq:hd_crit_avg_k2_user}\\
			&\qquad\times \left[ 1 - \left(d \log \left(\frac{\pi}{2}\right) - \dfrac{2}{\gamma-2}\right) \left(W_{-1}\left(-{\nu \over n}\right)\right)^{-1}\right].\nonumber
		\end{align}
		When $\gamma = 2$ in the critical regime, one sets
		\begin{subequations}
			\begin{align}
				\R &= {\rm ln} \left( n/ \nu\right),\\
				\m &= \R-  2 \ln \R,
			\end{align}
		\end{subequations}
		while $\nu$ is obtained by solving Eq.~(\ref{eq:hd_crit_gamma2_avg_k}), now taking the form of
		\begin{align}
			\langle k \rangle  & \approx  \nu { 2^{d} \over d I_{d,1}  } \frac{1}{\left[\ln\left(n / \nu \right)\right]^{2}} \nonumber\\
			&\qquad\quad\times  \left[   2{\rm Li}_{3}\left[-  \left[\ln \left (\frac{n} { \nu} \right)\right]^{2} \left(\frac{\pi}{2}\right)^{d} \right]\right. \nonumber\\
			&\qquad\qquad\quad - {\rm Li}_{3}\left[- \left[\ln \left (\frac{n} { \nu} \right)\right]^{2}  \left(\frac{\pi}{2}\right)^{d} \left({\nu \over n}\right) \right]\nonumber\\
			&\qquad\qquad\quad\left. - \, {\rm Li}_{3}\left[- \left[\ln \left (\frac{n} { \nu} \right)\right]^{2}  \left(\frac{\pi}{2}\right)^{d} \left({n \over \nu}\right) \right] \right].\label{eq:hd_crit_gamma2_avg_k2}
		\end{align}
	
	\subsection{Hot regime, $\tau>1$}
	\label{sec:hot}
	In the $\tau > 1$ case the Eqs.~(\ref{eq:hd_avgk2}) and  (\ref{eq:hd_avgk3}) can be approximated as
	\begin{eqnarray}
		\langle k \rangle_n & \approx & (n-1) \, \mathcal{I}(d,\tau) \, e^{\m/\tau} \langle e^{-\rr /\tau}\rangle^{2},\label{eq:hd_hot_avg_k}\\
		\langle k (\rr) \rangle_n & \approx & { \langle k \rangle \over \langle e^{-\rr/\tau}\rangle } e^{-\rr/\tau},
	\end{eqnarray}
	where 
	\begin{equation}
		\mathcal{I}(d,\tau) \equiv {\frac{1}{I_{d,1} } \int_{0}^{\pi}\frac {{\rm sin}^{d-1} \theta_1 {\rm d} \theta_1}{{\rm sin} \left(\frac{\theta}{2}\right)^{d/\tau}}}, 
		\label{eq:idtau}
	\end{equation}
	and
	\begin{equation}
		\langle e^{-\rr/\tau} \rangle \equiv \int_{0}^{\R} \rho_{\rr}(\rr) e^{-{\rr / \tau}} {\rm d} \rr \approx
		{\al \tau \over \al \tau -1} \left(e^{-\R / \tau} - e^{-\al/ \R}\right). 
		\label{eq:hd_avg_e-rt}
	\end{equation}
	Note that the expression for $\langle e^{-\rr/\tau}\rangle$ given by Eq.~(\ref{eq:hd_avg_e-rt}) is
	valid for all values of $\al \geq 1$ and $\tau > 1$ since $\al \tau >1$.

	Similar to the $\tau<1$ regime, we demand $\langle k_{max} \rangle_n \sim n$ and $\langle k_{min} \rangle_n \sim 1$ to
	obtain the scaling relationships for $\m$ and $\R$:
	\begin{equation}
		\m = \R = \tau \ln \left(n/\nu\right). 
		\label{eq:hd_hot}
	\end{equation}
	
\begin{figure}
		\includegraphics[width=3in]{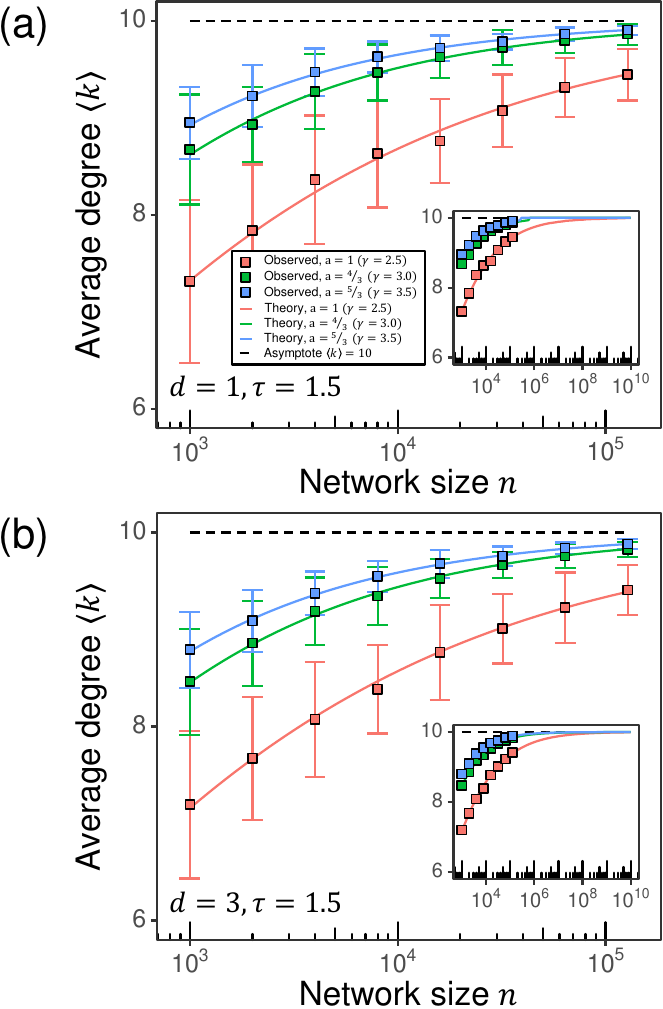}
		\caption{ \footnotesize Expected degree $\langle k \rangle$ as a function of network size $n$ for RHGs in the hot ($\tau = 1.5$) regime with (a) $d=1$ and (b) $d=3$. Each panel includes the results for (red) $\al=1$ ($\gamma = 2.5$),  (green) $\al=\frac{4}{3}$ ($\gamma = 3.0$), and (blue) $\al=\frac{5}{3}$ ($\gamma = 3.5$). Each point is the average of $100$ simulations and the error bars display standard deviations. Solid lines are theoretical values for $\langle k \rangle$ prescribed by Eqs.~(\ref{eq:hd_avgk2}) and~(\ref{eq:hd_avgk3}) and the dashed line is the thermodynamic limit for $\langle k \rangle$ given by Eq.~(\ref{eq:hd_hot_avg_k2}). The insets in (a) and (b) correspond to extended domains of $n$ values.}
		\label{fig:avgkhot}
	\end{figure}

	This scaling in combination with Eq.~(\ref{eq:hd_hot_avg_k}) leads in the thermodynamic limit to
	\begin{subequations}
		\begin{align}
			\langle k \rangle_n  & \approx  \nu \, \mathcal{I}(d,\tau) \left(\al \tau  \over \al \tau -1 \right)^{2},\label{eq:hd_hot_avg_k2}\\
			\langle k (\rr) \rangle_n  & \approx {n \over \nu} \left(\al \tau - 1 \over \al \tau \right)\langle k \rangle e^{-\rr/\tau},\label{eq:hd_hot_avg_kr}\\
			P(k) & \approx \al \tau \left[ \langle k (\R) \rangle \right] ^{\al \tau} \frac{\Gamma[k-\al \tau]}{\Gamma[k+1]} \sim k^{-\al \tau -1},\label{eq:hd_hot_pk}
		\end{align}
	\end{subequations}
	confirmed by Fig.~\ref{fig:avgkhot}, and Fig.~\ref{fig:kr_plots}(g-h).
	
\begin{figure}
		\includegraphics[width=3in]{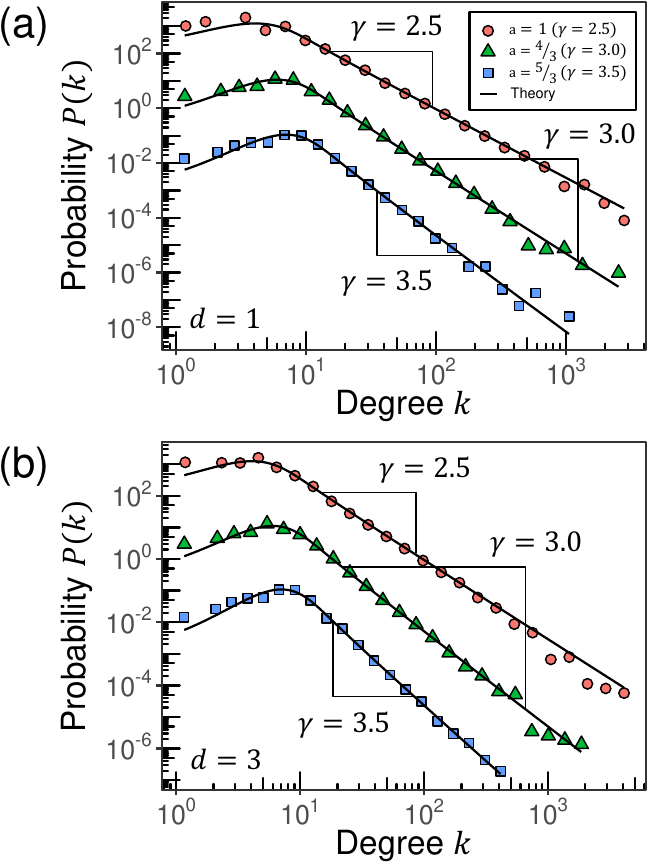}
		\caption{\footnotesize Degree distribution $P(k)$ for RHGs with (a) $d=1$ and (b) $d=3$ at $\tau = 1.5$ and $n = 1000 \cdot 2^{7}$. Each panel includes the degree distributions for (red) $\al=1$ ($\gamma = 2.5$),  (green) $\al=\frac{4}{3}$ ($\gamma = 3.0$), and (blue)  $\al=\frac{5}{3}$ ($\gamma = 3.5$). Degree distributions are binned logarithmically to suppress noise at large $k$ values. Solid lines are theoretical values for $P(k)$ prescribed by Eq.~(\ref{eq:hd_hot_pk}). For visibility, the probabilities corresponding to $\gamma = 3.0$ and $\gamma = 2.5$ are multiplied by $10^{2}$ and $10^{4}$, respectively. The scaling constant $\nu$ is chosen such that $ \nu = 10 \times {1 \over \mathcal{I}(d,\tau)} \left({\al \tau - 1\over \al \tau}\right)^{2}$, corresponding to $\langle k \rangle~=~10$ in the thermodynamic limit.}
		\label{fig:pk_hot}
	\end{figure}

	Similar to the cold and critical regimes, RHGs in the hot regime are sparse and have power-law degree distributions, $P(k) \sim k^{-\gamma}$. Different from the cold and critical regimes, the degree distribution exponent $\gamma = \al \tau + 1$ depends on both $\al$ and $\tau$ in the hot regime, as confirmed by Fig.~\ref{fig:pk_hot}.
	
\begin{figure}
		\includegraphics[width=3in]{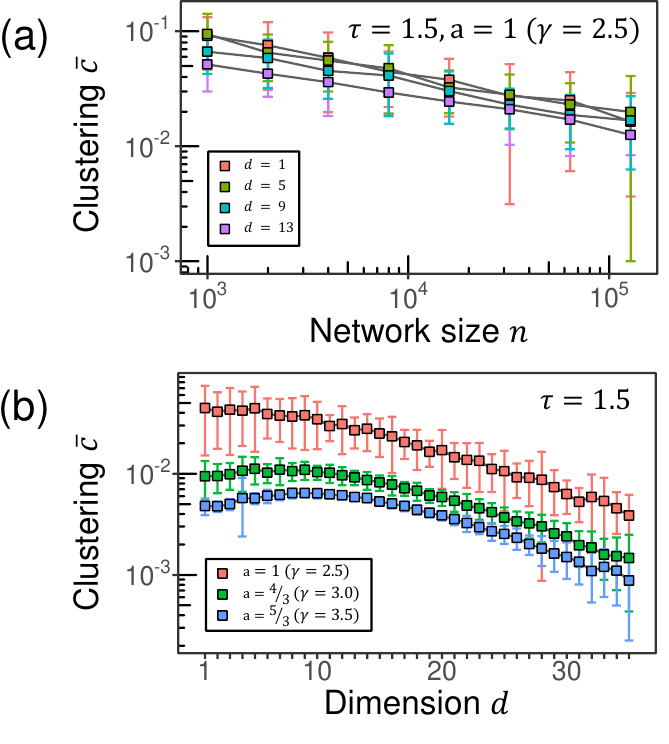}
		\caption{ \footnotesize Average clustering coefficient $\bar{c}$ (a) of nodes with degree $k > 1$ as a function of network size $n$ and (b) as a function of dimension $d$  at $\tau = 1.5$. Panel (a) includes results for (red) $d = 1$, (green) $d = 5$, (blue) $d = 9$ and (purple) $d = 13$, for $\al = 1$ ($\gamma = 2.5$), while panel (b) includes results for (red) $\al=1$ ($\gamma = 2.5$), (green) $\al=\frac{4}{3}$ ($\gamma = 3.0$), and (blue) $\al=\frac{5}{3}$ ($\gamma = 3.5$), for $n = 10^{4}$. Each point is the average of $100$ simulations and the error bars display standard deviations. The scaling constant $\nu$ is chosen such that $ \nu = 10 \times {1 \over \mathcal{I}(d,\tau)} \left({\al \tau - 1\over \al \tau}\right)^{2}$, corresponding to $\langle k \rangle~=~10$ in the thermodynamic limit. }
		\label{fig:hot_clust}
	\end{figure}

	In the hot regime, we observe that average clustering decays with $n$ as $n^{-\sigma}$, where the value of the exponent~$\sigma$ depends on both $d$ and $\tau$, see Fig.~\ref{fig:hot_clust}. It is already known~\cite{vanderkolk2022anomalous} that the scaling of average clustering with $n$ depends on $\tau$ in the $d=1$ case. Here, we observe that it also depends on the dimension $d \geq 1$. As discussed in Sec.~\ref{sec:cold}, the angular distance distribution between two random points on a $d$-sphere at $d \rightarrow\infty$ approaches the Dirac delta function centered at~$\pi/2$~\cite{hammersley1950}, causing clustering to vanish in the large-$d$ limit. Fig.~\ref{fig:hot_clust}(b) confirms that clustering in the hot regime decreases slowly with dimension~$d$.
	
		\subsubsection*{Graph property perspective, $\tau > 1$}
		To generate an RHG in the hot regime with given average degree $\langle k \rangle$ and power-law exponent $\gamma > 2$, one needs to set
		\begin{subequations}
			\begin{align}
				\al &= \frac{\gamma -1}{\tau},\\
				\m &= \R = \tau \ln \left( n / \nu \right),
			\end{align}
		\end{subequations}
		where $\nu$ is given by 
		\begin{equation}
			\langle k \rangle  \approx   \nu  \,\mathcal{I}(d,\tau) \, \left(\gamma  - 1 \over \gamma -2 \right)^{2} ,\\
			\label{eq:hd_hot_avg_k2_user}
		\end{equation}
		and $\mathcal{I}(d,\tau)$ is given by Eq.~(\ref{eq:idtau}).
		
		Because we require $\al > 1$, in this regime the power-law exponent is bounded $\gamma - 1 > \tau$. Since $\tau > 1$, a power-law exponent $\gamma = 2$ is not possible in the hot regime.
		
\begin{figure*}[h]
		\includegraphics[width=0.99\textwidth]{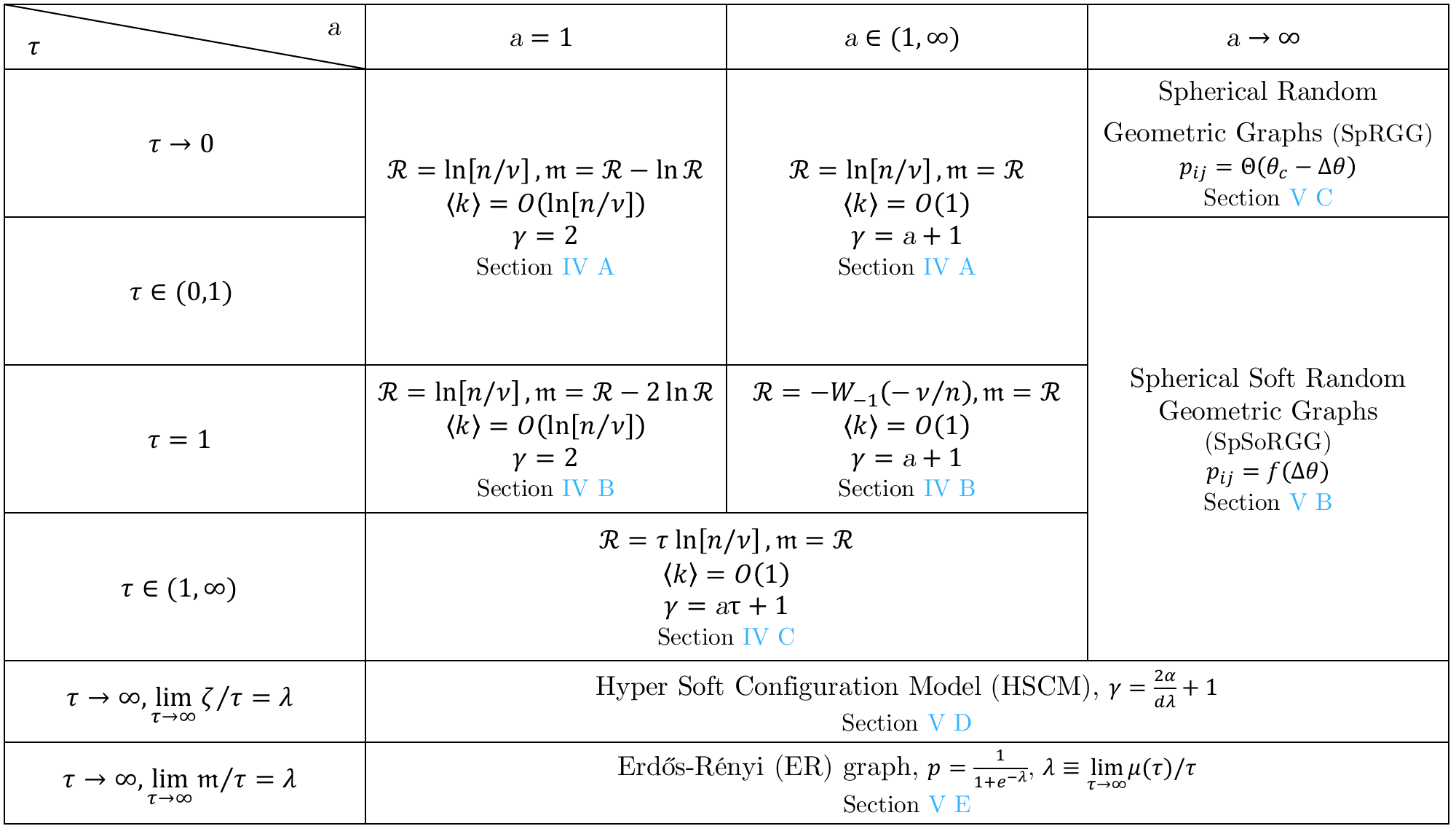}
		\caption{ \footnotesize RHG regimes in terms of rescaled variables.} 
		\label{fig:table_full}
	\end{figure*}

\begin{figure*}[h]
		\includegraphics[width=0.97\textwidth]{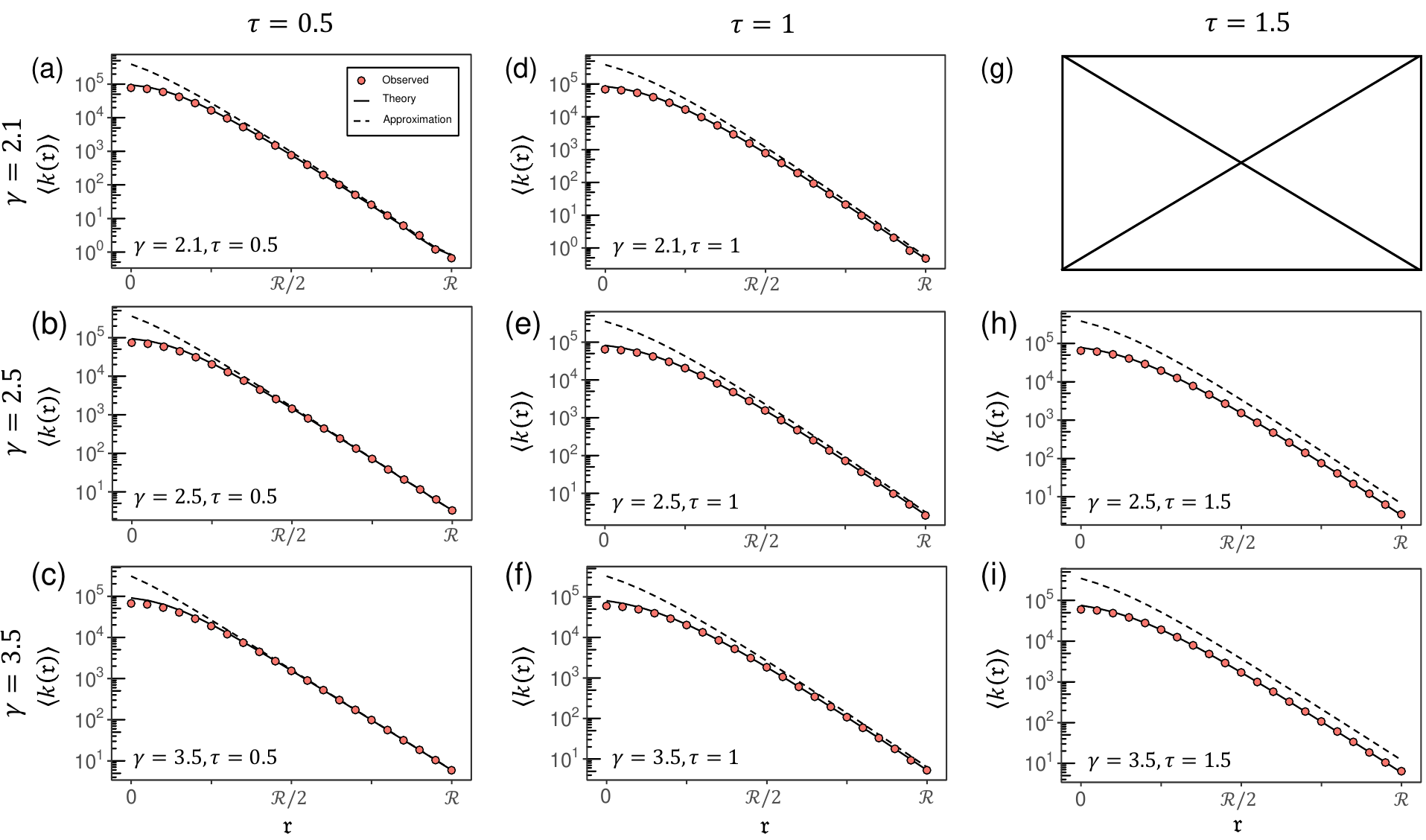}
		\caption{ \footnotesize Average degree $\langle k(\rr) \rangle$ for $\rr \in [0,\R]$ and $n=10^{5}$, $d=3$ and $\langle k\rangle=10$. Panels correspond to (a-c) the cold regime at $\tau = 0.5$, (d-f) the critical regime at $\tau = 1$, and (g-i) the hot regime at $\tau = 1.5$. In panels (a,d,g) $\gamma = 2.1$, (b,e,h) $\gamma = 2.5$, and (c,f,i) $\gamma = 3.5$. The combination $\gamma=2.1$, $\tau = 1.5$ in panel (g) implies $\al<1$, which is not possible in our framework. Each point is the average of 100 simulations. Solid lines are theoretical values for $\langle k(\rr) \rangle$ prescribed by Eq.~(\ref{eq:hd_avgk2}), and dashed lines are the same values but with approximations $\sin x \approx x$.}
		\label{fig:kr_plots}
	\end{figure*}

\begin{figure}
		\centering
		\includegraphics[width=3in]{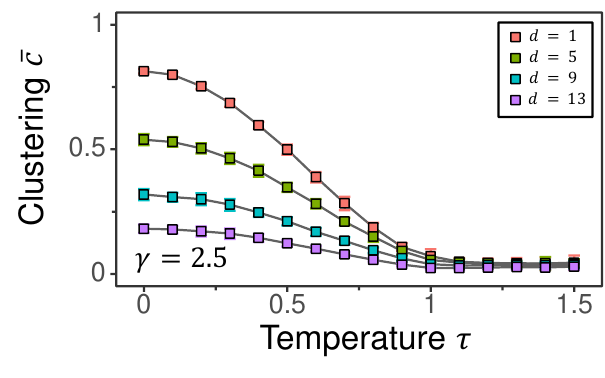}
		\caption{ \footnotesize Average clustering coefficient $\bar{c}$ of nodes with degree $k > 1$ as a function of temperature $\tau$ for different dimensions $d$ and $n = 10^{4}$. Each point is the average of $100$ simulations and the error bars display standard deviations (not visible for most points here). The scaling constant $\nu$ is chosen such that $\langle k \rangle = 10$ in the thermodynamic limit.}
		\label{fig:clust}
	\end{figure}

		In summary, we find that under the proper change of variables prescribed by Eq.~(\ref{eq:rescaling}), RHGs of any dimensionality can be naturally described by three regimes based on the rescaled temperature $\tau$, cold regime ($0 \leq \tau < 1$), critical regime ($\tau = 1$) and hot regime ($\tau > 1$). In each of these regimes, RHGs of any dimensionality $d \geq 1$ exhibit similar topological properties with respect to node degrees, Fig.~\ref{fig:table_full}. Our approximations of $\langle k (\rr) \rangle$ work well in each of the three regimes, Fig.~\ref{fig:kr_plots}, but there is a small constant bias in the hot regime. The approximations break down for nodes close to the center of $\mathbb{B}^{d+1}$ (values of $\rr$ close to $0$) across all three regimes. Consistent with the $d = 1$ case, Ref.~\cite{Krioukov2010hyperbolic}, the average clustering coefficient in RHGs seems to approach a constant in the thermodynamic limit when $\tau < 1$, decreases polynomially as the RHG size increases when $\tau > 1$, and decreases logarithmically as a function of network size in the critical regime when $\tau = 1$. We find that the average clustering coefficient in the RHG model decreases in the $d\rightarrow\infty$ limit in each of the cold, critical, and hot regimes, consistent with findings for the GIRG model. Finally, Fig.~\ref{fig:clust} shows that for any dimension, clustering is a decreasing function of~$\tau$.
	
	\section{Limiting cases of the RHG model}
	\label{sec:limiting}
	In this section, we analyze several important parameter limits of the RHG and show that they correspond to well-known graph ensembles.
	
	\subsection{$\tau \to 0$ limit in the cold regime}
	The case of $\tau=0$ is well-defined as the $\tau \to 0$ limit of the cold regime.
	The $T\to0$ limit of the connection probability function in Eq.~(\ref{eq:conn_hypergeometric_ERG}) is the step function
	\begin{equation}
		p_{ij} = \Theta\left(\mu - d_{ij} \right),
	\end{equation}
	such that connections are established deterministically between node pairs separated by distances
	smaller than~$\mu$.

	In this case we have
	${\pi \tau \over \sin\left( \pi \tau\right)} \to 1$ in (\ref{eq:hd_cold_avg_k2}), leading to
	\begin{align}
		\langle k  \rangle_n & \approx { \nu 2^{d} \over d I_{d,1}} \left({\al \over \al - 1}\right)^{2}\nonumber\\
		&\qquad\times \left[1 - 2 \left(\frac{n}{\nu}\right)^{1-\al} + \left(\frac{n}{\nu}\right)^{2(1-\al)}\right],\\
		\langle k (\rr) \rangle_n & \approx {n \over \nu} \left({\al -1 \over \al}\right)
		\langle k  \rangle e^{-\rr} \sum_{\ell\,=\,0}^{\infty}  \left(\frac{\nu}{n}\right)^{\ell (\al-1)}.
	\end{align}
	The resulting graphs are sparse and are characterized by a power-law degree distribution $P(k) \sim k^{-\gamma}$, $\gamma = \al + 1$, similar to the $0 <\tau < 1$ case.

	\subsection{$\al \to \infty$ limit: Spherical Soft Random Geometric Graphs (SpSoRGG)}
	\label{sec:spsrgg}
	
	In this limit the radial coordinate distribution (\ref{eq:hd_rho_distr_rescaled}) degenerates to
	\begin{equation}
		\rho_{\rr}(\rr) \to \delta\left( \rr - \R \right).
	\end{equation}
	As a result, all nodes are placed at the boundary of the hyperbolic ball $\mathbb{B}^{d+1}$ with $\rr_{i} = \R$. Even though the
	distances between nodes are still hyperbolic, they are fully determined by the angles on $\mathbb{S}^{d}$,
	\begin{equation}
		\zeta d_{ij}=  \cosh^{-1}\left[ \cosh \left(\frac{2\R}{d}\right)^{2} - \sinh \left(\frac{2\R}{d}\right)^{2}  \sin \left(\Delta\theta_{ij}
		\right)\right].\nonumber
	\end{equation}
	Hence, the connection probabilities $\{p_{ij}\}$ are fully determined by angles $\Delta\theta_{ij}$,
	\begin{equation}
		p_{ij} = {1 \over 1 + \exp\left( { \tilde{d}\left(\Delta\theta_{ij} \right) - \m \over \tau}\right)},
	\end{equation}
	where $\tilde{d} \left(\Delta\theta_{ij} \right)\equiv  \frac{d \zeta}{2} d_{ij}$.

	In the $\al \to \infty$ regime, all nodes are effectively placed at the surface of the unit sphere $\mathbb{S}^{d}$, and connections are made with distance-dependent probabilities on the sphere. Hence, in the
	$\al \to \infty$ limit, RHGs are soft RGGs on $\mathbb{S}^{d}$.

	\subsection{$\al \to \infty$, $\tau \to 0$ limit: Spherical Random Geometric Graphs (SpRGG)}
	\label{sec:sprgg}
	
	If $\al \to \infty$ and $\tau \to 0$, the connection probabilities in Eq.~(\ref{eq:conn_rescaled_ERG}) become 
	\begin{equation}
		p_{ij} = \Theta (\theta_{c} - \Delta\theta_{ij}),
	\end{equation}
	where $\theta_{c}$ is the solution to the equation $\tilde{d}(\theta_c) = \m$. Thus, in this limit the RHG becomes the sharp random geometric graph on $\mathbb{S}^{d}$ (SpRGG).

	The expected degree of the SpRGG equals the expected number of nodes that fall within an angle $\theta_{c}$ of the $\theta_1=0,...,\theta_{d}=0$ point,
	\begin{equation}
		\langle k \rangle = (n-1)\,\tilde{p},
		\label{eq:avgk_sprgg}
	\end{equation}
	where the volume of the $(d-1)$-dimensional sphere of radius $\theta_{c}$ in $\mathbb{S}^{d}$ is 
	\begin{equation}
		\tilde{p} = \frac{\int_{0}^{\theta_{c}} \left[{\rm sin}\left( \theta\right)\right]^{d-1} {\rm d} \theta  }{\int_{0}^{\pi} \left[{\rm sin}\left( \theta\right)\right]^{d-1} {\rm d} \theta  }.
	\end{equation}
	The degree distribution is thus binomial,
	\begin{equation}
		P(k) = {\rm Bin}\left[ n-1, \tilde{p} \right](k),
	\end{equation}
	converging to the Poisson distribution with mean~$\langle k \rangle$ if $\theta_c$ is such that $n\tilde{p}\to\langle k \rangle$. Since the Poisson distribution is the $\gamma\to\infty$ limit of the Pareto-mixed Poisson distribution~\eqref{eq:pk_plaw}, we refer to this regime as the $\gamma\to\infty$ case in Fig.~\ref{fig:table_full}.

	\subsection{$\zeta \to \infty$, $\tau \to \infty$ limit: Hyper Soft Configuration Model (HSCM)}
	\label{sec:hscm}

	In the $\zeta \to \infty$ limit, the hyperbolic distances in (\ref{eq:cosh-law}) degenerate to
	\begin{equation}
		d_{ij} = r_{i} + r_{j},
	\end{equation}
	such that the angular coordinates of nodes are ignored in this limit. Further, if $\tau$ also tends to infinity, $\tau \rightarrow \infty$, but such that $\lim_{\zeta \to \infty} \frac{\zeta}{\tau} = \lambda > 0$,  where $\lambda$ is a constant, then the connection probability in Eq.~(\ref{eq:conn_hypergeometric_ERG}) simplifies to
	\begin{equation}
		p_{ij}= {1 \over 1 + e^{\omega_i}e^{\omega_j}},
	\end{equation}
	which is the connection probability in the Hyper Soft Configuration Model (HSCM)~\cite{VanderHoorn2018sparse}. Here, $\omega_{i}~=~\frac{d\lambda}{2}\left(r_i- \frac{\mu}{2}\right)$ are the Lagrange multipliers controlling expected node degrees.
	The Lagrange multipliers are drawn from the effective pdf
	\begin{align}
		\rho_{\omega}(\omega) & \approx \frac{2\alpha}{d \lambda}e^{\alpha\left(\frac{\mu}{2} - R\right)} e^{\frac{2\alpha}{d \lambda} \omega},\\
		\omega &\in \left(-\frac{d\lambda \mu}{4}, \frac{d\lambda}{2} \left(R - \frac{\mu}{2}\right)\right).
	\end{align}

	The expected degrees in the HSCM are approximated by
	\begin{align}
		\langle k (\omega_i) \rangle & \approx  (n-1) \langle e^{-\omega} \rangle e^{-\omega_i}, \\
		\langle k \rangle & \approx  (n-1) \langle e^{-\omega} \rangle^{2},
	\end{align}
	where $\langle e^{-\omega} \rangle \equiv \int {\rm d} \omega \rho_{\omega}(\omega) e^{-\omega}$. By demanding that $\langle k \left( \omega (r=0) \right) \rangle_n \sim n$ and $\langle k \left( \omega (r= R) \right	) \rangle_n \sim 1$ we
	obtain $R = \frac{2}{d\lambda} \ln (n)$, while  $\mu  = R$ in the case  of $\frac{2 \alpha }{d \lambda} > 1$, and $\mu= \frac{2\alpha}{d \lambda}  R$ in the case of $\frac{2 \alpha }{d	\lambda} < 1$.

	In both cases, $\langle k (\omega)\rangle \sim e^{-\omega}$ and graphs in the HSCM are sparse, while the conditional probability $P(k|\omega)$ is well-approximated by the Poisson distribution:
	\begin{equation}
		P(k|\omega) \approx {1 \over k!} e^{-\langle k (\omega)\rangle } \left[\langle k (\omega)\rangle \right]^{k},\label{eq:hscm_cond_pk}
	\end{equation}
	see Ref.~\cite{VanderHoorn2018sparse}. The resulting degree distribution $P(k)$ is a mixed Poisson distribution:
	\begin{equation}
		P(k) \approx  {1 \over k!}  \int_{-\frac{d\lambda \mu}{4}}^{\frac{d\lambda}{2} \left(R - \frac{\mu}{2}\right)}  e^{-\langle k (\omega)\rangle }   \left[\langle k (\omega)\rangle \right]^{k} \rho_{\omega}(\omega) {\rm d} \omega, \label{eq:hscm_pk}
	\end{equation}
	with mixing parameter $\langle k (\omega)\rangle$. Using (\ref{eq:hd_cond_pk}) and (\ref{eq:hd_pk}) we obtain
	\begin{equation}
		P(k) \approx \left(\gamma -1 \right)  \kappa_{0}^{\gamma - 1} {\Gamma[k+1- \gamma , e^{\frac{d\lambda}{2} (R-\frac{\mu}{2})} \kappa_{0}] \over \Gamma[k + 1]} \sim k^{-\gamma},\label{eq:hscm_pk_plaw}
	\end{equation}
	where $\gamma = {2\alpha \over d\lambda} + 1$ and $\kappa_{0} \equiv e^{\frac{d\lambda}{2}\left(\frac{\mu}{2} - R\right)} {\langle k \rangle \over \langle e^{-\omega} \rangle}$.
	
	Thus, the RHG model in the $\zeta \to \infty$, $\tau \to \infty$, $\zeta/\tau\to\lambda$ limit degenerates to the HSCM with a scale-free degree distribution with exponent $\gamma = {2\alpha \over d\lambda} + 1$.
	
	\subsection{$\tau\to \infty$ limit: the Erd\H{o}s-R\'{e}nyi (ER) graph}
	\label{sec:er}
	
	The limit of $\tau \to \infty$ and finite $\zeta$ is the most degenerate case. Indeed, in this regime connection probabilities $p_{ij}$ become independent of the hyperbolic distances $d_{ij}$ between the nodes:
	\begin{equation}
		\label{eq:ER_mode}
		p_{ij} = \lim_{\tau \to \infty} \frac{1}{1 + e^{-\mu(\tau)/\tau}}.
	\end{equation}
	It is seen from Eq.~(\ref{eq:ER_mode}) that the connection probabilities are non-trivial only in the case $\mu(\tau) \sim \tau$. In this case, the connection probabilities are constant,
	\begin{equation}
		\label{eq:ER_mode2}
		p_{ij} = p = \lim_{\tau \to \infty} \frac{1}{1 + e^{-\lambda}},
	\end{equation}
	where $\lambda \equiv \lim_{\tau \to \infty} \frac{\mu(\tau)}{\tau}$. By varying $\lambda \in (-\infty, \infty)$ one can tune the connection probabilities $p \in (0,1)$ of the resulting ER graphs.
	
	One can also check that the ER limit can be obtained either as the $\gamma\to\infty$ ($\alpha\to\infty$ or $\lambda\to0$) limit of the HSCM, or as $\tau\to\infty$ limit of SpSoRGGs.

		\subsection{$d \to \infty$ limit.} \label{sec:dinf}
		
		As dimension $d$ increases, the angular distance distribution between two random points on a $d$-sphere approaches the Dirac delta function centered at~$\pi/2$~\cite{hammersley1950}. As a result, the role of a node's angular coordinates in the hyperbolic distances diminishes, and the network becomes more similar to the HSCM. The distances between nodes depend only on their radial coordinates,
		\begin{equation}
			d_{ij} \approx \rr_{i} + \rr_{j} - \frac{1}{\zeta} \ln 2.
		\end{equation}

\begin{figure}
			\includegraphics[width=3in]{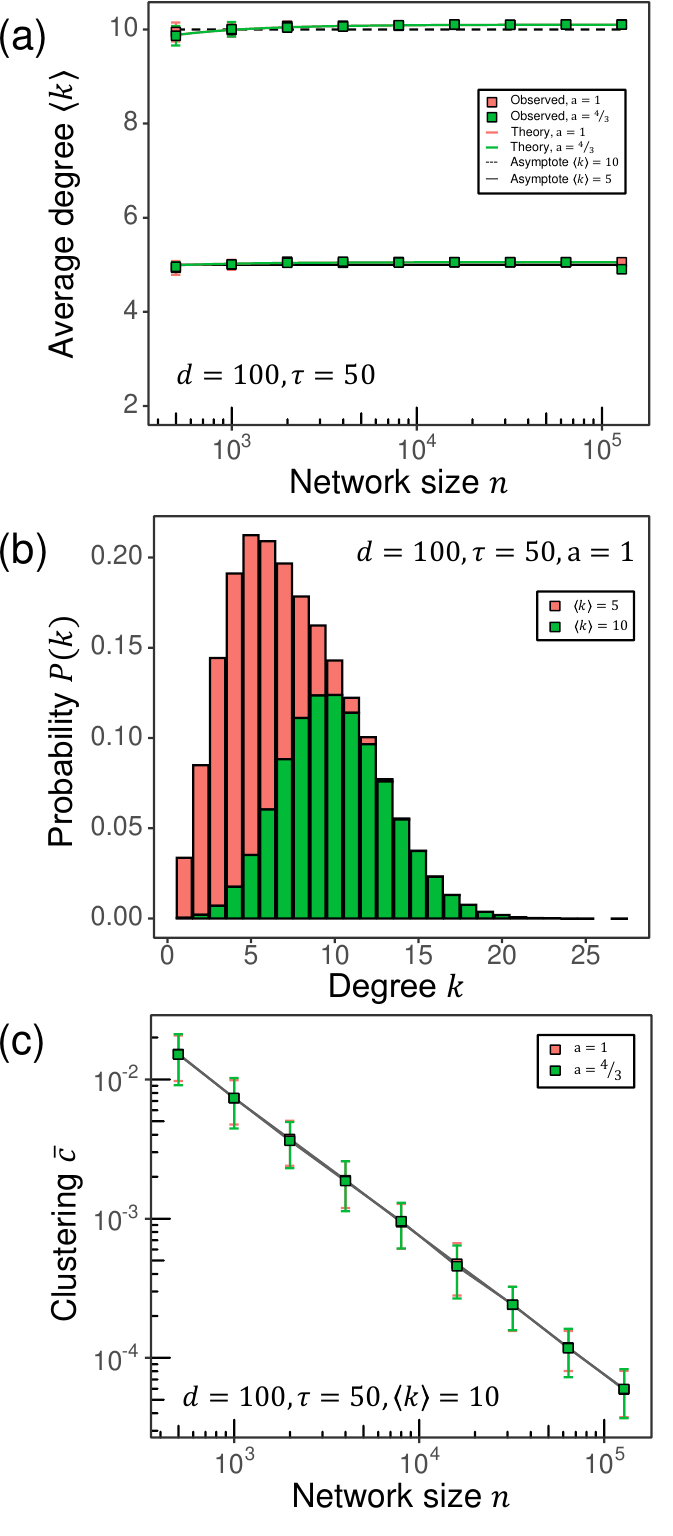}
			\caption{ \footnotesize{ Expected degree $\langle k \rangle$ (a) as a function of network size~$n$, degree distribution~$P(k)$ (b), and average clustering coefficient~$\bar{c}$ (c) of nodes with degree $k > 1$ as a function of network size~$n$ for RHGs with $d=100$ at $\tau = 50$. Panels (a) and (c) include results for (red) $\al = 1$ and (green) $\al = 4/3$, panel (b) includes results for (red) $\langle k \rangle = 5$ and (green) $\langle k \rangle = 10$ at $\al = 1$ and $n = 1000 \cdot 2^{7}$, while $\langle k \rangle = 10$ in panel (c). Each point in panels (a) and (c) is the average of $20$ simulations and the error bars display standard deviations, while probabilities $P(k)$ in panel (b) are obtained from a single network realization for each different value of $\langle k \rangle$. The scaling constant $\nu$ is chosen such that $\nu = \langle k \rangle\, 2^{-\frac{1}{2T}}$.
			}}
			\label{fig:dinf}
		\end{figure}

		Using this approximation, we estimate the expected degree of a node at $\rr$ in the $d \gg 1$ regime as
		\begin{equation}
			\langle k \left( \rr \right)\rangle \approx (n-1) \, e^{- \al \R} \int_{1}^{e^{\al \R}} \frac{{\rm d} \xi}{1 + \left(\frac{1}{2}\right)^{\frac{d}{2\tau}} e^{\frac{\rr - \m}{\tau}} \xi^{\frac{1}{\al \tau}}  }.
		\end{equation}
		Since $\tau \to \infty$ as $d\to \infty$ for any positive $T$, we are in the hot regime, $\tau > 1$. In this regime, we set $\m = \R = \tau \ln \left(\frac{n}{\nu}\right)$, resulting in 
		\begin{subequations}
			\begin{eqnarray}
				\langle k\left( \rr\right)\rangle \rangle_{n} &\approx& n \,2^{\frac{d}{2 \tau}} e^{-\frac{\rr}{\tau}},\\
				\langle k \rangle_{n} &\approx& \nu \,2^{\frac{d}{2 \tau}}
			\end{eqnarray}
		\end{subequations}
		Thus, high-dimensional RHGs are akin to low-dimensional RHGs in the hot regime. Indeed, as seen from Fig.~\ref{fig:dinf}(a), high-dimensional RHGs are sparse graphs.  
		
		High-dimensional RHGs are well-defined in the $d \to \infty$~limit. In this case, RHGs are sparse graphs with $\langle k \rangle_{n} \approx \nu \, 2^{\frac{1}{2 T}}$. In the $d \to \infty$ limit, $\gamma \to \infty$ implies that RHGs are no longer described by a power-law degree distribution. This is indeed the case, as all nodes are located at the boundary of $\mathbb{B}^{d+1}$, $\rr = \R$, leading to a Poisson degree distribution,
		\begin{equation}
			P(k) = P(k| \rr = R) \approx \frac{1}{k!} e^{-\langle k \rangle} \langle k \rangle^{k},
		\end{equation}
		where $ \langle k \rangle = \langle k(\rr = \R) \rangle \approx \nu \, 2^{\frac{1}{2T}}$,
		see Fig.~\ref{fig:dinf}(b). 
		
		The average clustering coefficient of high-dimensional RHGs decreases as a function of network size, similar to RHGs of finite dimensionality in the hot regime, Fig.~\ref{fig:dinf}(c). 
		
		\subsubsection*{Graph property perspective, limiting cases}
\begin{figure}
		\includegraphics[width=3in]{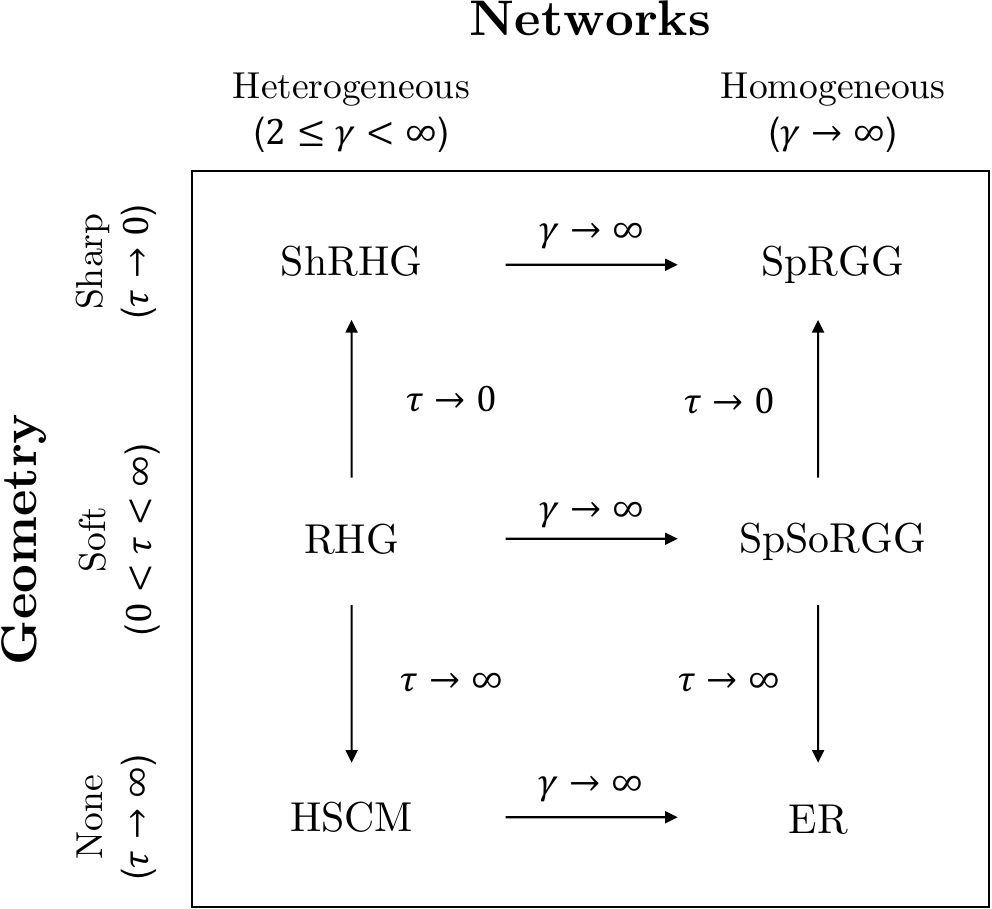}
		\caption{ \footnotesize Limiting regimes of the RHG from the graph property perspective.} 
		\label{fig:user_persp}
	\end{figure}

		Figure~\ref{fig:user_persp} summarizes the properties of the RHG and its limiting cases in the $(\tau, \gamma)$ phase space. Within the $(\tau, \gamma)$ phase space, all the RHG temperature regimes condense into the heterogeneous ($2\leq\gamma<\infty$) soft-geometric ($0\leq\tau<\infty$) state.

		The sharp-geometric limit ($\tau \to 0$) of this state is well-defined and is obtained by taking the $\tau \to 0$ limit in Eq.~(\ref{eq:hd_cold_avg_k2_user}). In this case, to generate RHGs with desired expected degree $\langle k \rangle$ and power-law distribution exponent $\gamma > 2$, one needs to set $\m = \R = \ln(n/\nu)$, where 
		$\nu$ is given by 
		\begin{equation}
			\label{eq:hd_cold_avg_k2_user_sharp}
			\langle k \rangle  \approx \nu  {2^{d} \over d I_{d,1}} \left({ \gamma -1 \over \gamma - 2}\right)^{2}  \left[1 - 2 \left(\frac{n}{\nu}\right)^{2-\gamma} + \left(\frac{n}{\nu}\right)^{2(2-\gamma)}\right].
		\end{equation}

		By setting $\gamma \to \infty$ ($\al \to \infty$) in the RHG, one arrives at Spherical Soft Random Geometric Graphs (SpSoRGG). Here nodes are placed at the  boundary of the $\mathbb{B}^{d+1}$ ball, and connections are established with probabilities dependent on distances between the nodes on its $\mathbb{S}^{d}$ boundary, see Sec.~\ref{sec:spsrgg}. Since SpSoRGGs are characterized by the Poisson degree distribution, we refer to them as the homogeneous ($\gamma \to \infty$) soft-geometric limit of the RHG. The expected degree of the SpSoRGG can be obtained by taking the $\gamma \to \infty$ limit of the RHG in the cold, critical, or hot regimes, depending of the $\tau$ value. In other words, to generate a SpSoRGG with prescribed $\tau$ and $\langle k \rangle$, one needs to set $\m$, $\R$, and $\nu$ as follows
		\begin{subequations}
			\begin{align}
				0 < \tau < 1&: \m = \R = \ln \left(n/\nu\right), ~\langle k \rangle  \approx  {\nu 2^{d} \over d I_{d,1}}{\pi \tau \over \sin(\pi \tau)} ;\\
				\tau = 1&: \m = \R = - W_{-1}(\nu/n),~\langle k \rangle  \approx   {\nu 2^{d} \over d I_{d,1}  };\\
				\tau > 1&: \m = \R = \tau \ln \left(n/\nu\right), ~ \langle k \rangle  \approx   \nu  \,\mathcal{I}(d,\tau).
			\end{align}
		\end{subequations}

		By taking the $\tau \to 0$ limit of the Spherical Soft RGG we arrive at the Spherical Sharp RGG, or simply Spherical Random Geometric Graph (SpRGG). Similar to its soft counterpart, nodes in the SpRGG are placed at the $\mathbb{B}^{d+1}$ boundary but connections are established deterministically between nodes separated by distances smaller than the threshold, Sec.~\ref{sec:spsrgg}. Another possibility to arrive at the SpRGG is by taking the $\gamma \to \infty$ limit of the Sharp RHG. One can generate Spherical Sharp RGGs with the desired expected degree $\langle k \rangle$ by setting
		$\m = \R = \ln \left(n/\nu\right)$, and selecting $\nu$  from
		\begin{equation}
			\langle k \rangle  \approx  {\nu 2^{d} \over d I_{d,1}}.
		\end{equation}

		While both the hyper soft configurational model (HSCM) and the Erd\H{o}s-R\'{e}nyi (ER) graph are the $\tau \to \infty$ limits of the RHG, they belong to two distinct classes, as seen from the graph property perspective.

		The HSCM belongs to the non-geometric ($\tau\to\infty$) heterogeneous ($2 \leq \gamma < \infty$)  case and is the $\tau \to \infty$, $\zeta \to \infty$ limit of the RHG. To build RHGs with desired expected degree $\langle k \rangle$ and a power-law degree distribution exponent $\gamma > 2$, one sets $\mu=R = \frac{2}{d \lambda} \ln \left(\frac{n}{\nu}\right)$, where $\nu$ is the solution of 
		\begin{equation}
			\langle k \rangle \approx \nu \left(\frac{\gamma-1}{\gamma-2}\right)^{2} \left[ 1- \left(\frac {\nu}{n}\right)^{\gamma-2} \right]^{2}.
		\end{equation}

		The ER graph, on the other hand, belongs to the non-geometric ($\tau \to \infty$) homogeneous ($\gamma \to \infty$) state and is a $\gamma \to \infty$ limit of the HSCM. Alternatively, the ER graph can also be attained as the $\tau \to \infty$, $\zeta \to \infty$ limit of the SpSoRGG. 
		
		In the $d \rightarrow \infty$ limit, RHGs have a Poisson degree distribution, and one can generate RHGs with the desired expected degree by selecting $\nu$ from 
		\begin{equation}
			\langle k \rangle \approx \nu \, 2^{\frac{1}{2T}}.
		\end{equation}

	\section{Hyperbolic graph generator in $d+1$ dimensions}
	\label{sec:code}

	We conclude by presenting a software package that generates RHGs of arbitrary dimensionality, to be specified by the user. The generator covers the cold \mbox{($\tau<1$)}, critical \mbox{($\tau=1$)} and hot \mbox{($\tau > 1$)} regimes. The software package and detailed instructions on how to compile and use it are available from the Bitbucket repository~\cite{budel2022}. 
	
	The RHG generator can operate in two different modes: hybrid and model-based. In hybrid mode, the user provides expected degree $\langle k \rangle$, power-law exponent~$\gamma$, rescaled temperature~$\tau$ and dimension~$d$. Equations~(\ref{eq:hd_avgk2}) and~(\ref{eq:hd_avgk3}) are solved for the rescaled radius~$\R$ that yields the desired $\langle k \rangle$ using the bisection method. The triple integral that is found by combining Eqs.~(\ref{eq:hd_avgk2}) and~(\ref{eq:hd_avgk3}) is evaluated numerically using Monte Carlo integration with importance sampling through the GNU Scientific Library (GSL) \cite{Galassi2018scientific}. In model-based mode, the user directly provides the model parameters $\al$, $\tau$, $\R$~(or~$\nu$) and~$d$. We expect the model-based mode to be instrumental for research purposes.\\

	\section{Summary}\label{sec:summary}
	In this work, we have generalized random hyperbolic graphs (RHGs) to arbitrary dimensions. In doing so, we have found the rescaling of network parameters given by Eq.~(\ref{eq:rescaling}) that allows for reducing RHGs of arbitrary dimensionality to a single mathematical framework. Summarized in Fig.~\ref{fig:table_full}, our results indicate that RHGs exhibit similar connectivity properties, regardless of their dimension $d$. At the same time, higher dimensional realizations of the RHG model differ from the original $d=1$ RHG model with respect to other structural properties.
	
	One such property is clustering. We find that the degree-dependent and average clustering coefficients behave differently depending on the temperature regime. In the cold regime, $0 \leq \tau < 1$, RHGs are characterized by nonvanishing average and degree-dependent clustering. In the hot phase, $\tau >1$, clustering becomes size-dependent and vanishes in the thermodynamic limit. These observations are expected and have been previously studied in the special $d=1$ case~\cite{candellero2016clustering,Krioukov2010hyperbolic,fountoulakis2021clustering}. The critical temperature of $\tau = 1$ corresponds to a continuous phase transition, which has been shown in $d=1$ to be topological in nature, characterized by diverging entropy and the atypical finite-size scaling behavior of clustering~\cite{vanderkolk2022anomalous}. RHGs of arbitrary dimensionality allow us to study the behavior of clustering as a function of the dimension. To this end, we observe that, in general, clustering decreases as a function of $d$ in all three regimes. This observation is consistent with the $d\to \infty$ limit, which is akin to the hypersoft configurational model (HSCM). 
	
	In general, we note that the degree-dependent clustering does depend on both dimensionality and temperature based on our numerical experiments. This observation is in line with another work proposing to use the density of cycles to estimate network dimensionality~\cite{almagro2022detecting}. Yet, it remains an open question what exactly is different between two RHGs of different dimensionalities whose clustering is matched by selecting appropriate temperatures.
	
	Higher-dimensional RHGs may be instrumental in graph embedding tasks. Indeed, the dimensionality of the latent space has been shown to impact the accuracy of many network inference tasks, including link prediction, clustering, and node classification~\cite{gu2021principled}. One of the standard mapping approaches is maximum likelihood estimation (MLE), finding node coordinates of the network of interest by maximizing the likelihood that the network was generated as an RHG in the latent space. The likelihood function in the case of $\mathbb{H}^{2}$ has been shown to be extremely non-convex with respect to node coordinates~\cite{Papadopoulos2015network1}, making standard learning tools, like stochastic gradient descent, inefficient. Raising the dimensionality of the latent space $\mathbb{H}^{d+1}$ may lift some of the local maxima of the likelihood function, potentially leading to faster and more accurate graph embedding algorithms~\cite{jankowski2023d}.
	
	\section{Acknowledgments}
	We thank F.~Papadopoulos, M.~\'{A}.~Serrano, M.~Bogu\~{n}\'{a}, P.~van der Hoorn, and T.~van der Zwan for useful discussions and suggestions. This work was supported by ARO Grant No.~W911NF-17-1-0491 and NSF Grants No.~IIS-1741355 and CCF-2311160. G.~B. was supported by the NExTWORKx project, a collaboration between TU~Delft and KPN on future telecommunication networks. M.K. acknowledges the Dutch Research Council (NWO) grant OCENW.M20.244.

	\bibliographystyle{prx-bibstyle}

\end{document}